\documentclass[nonacm]{acmart}

\AtBeginDocument{%
  \providecommand\BibTeX{{%
    \normalfont B\kern-0.5em{\scshape i\kern-0.25em b}\kern-0.8em\TeX}}}

\setcopyright{none}
\renewcommand\footnotetextcopyrightpermission[1]{} 
\setcopyright{none}
\copyrightyear{2024}
\acmYear{2024}
\acmDOI{XXXXXXX.XXXXXXX}

\acmConference[]{}{}{}

\usepackage{xcolor}
\usepackage{caption}
\usepackage{subcaption}
\usepackage{longtable}
\usepackage{float}
\usepackage{pdfpages}
\usepackage{tikz}
\usepackage{xspace}
\usepackage{ifthen}
\usepackage{hyperref}

\usepackage{colortbl}
\usepackage{tabularx}
\usepackage{graphicx}

\usepackage{enumitem}
\usepackage[official]{eurosym}

\usepackage [english]{babel}
\usepackage [autostyle, english = american]{csquotes}
\MakeOuterQuote{"}

\usepackage{ifthen}
\newboolean{include-notes}
\setboolean{include-notes}{true}
\newcommand{\nb}[3]{\ifthenelse{\boolean{include-notes}}{{\colorbox{#2}{\bfseries\sffamily\scriptsize\textcolor{white}{#1}}}{\ \textcolor{#2}{\sf\small\textit{#3}}}}{}}

\newcommand{\amsalgorithm}{\textit{AMS algorithm}\xspace}

\makeatletter
\def\NAT@spacechar{~}
\makeatother

\begin{document}


\title[Information That Matters: Exploring Information Needs of People Affected by Algorithmic Decisions]{Information That Matters: Exploring Information Needs of People Affected by Algorithmic Decisions}

\author{Timothée Schmude}
\email{timothee.schmude@univie.ac.at}
\affiliation{%
  \institution{University of Vienna, Faculty of Computer Science, Research Network Data Science, Doctoral School Computer Science}
  \streetaddress{Währinger Straße 29}
  \city{Vienna}
  \state{Vienna}
  \country{Austria}
  \postcode{1090}
}

\author{Laura Koesten}
\email{laura.koesten@univie.ac.at}
\affiliation{%
  \institution{University of Vienna, Faculty of Computer Science, Research Group Visualization and Data Analysis}
  \streetaddress{Sensengasse 6}
  \city{Vienna}
  \state{Vienna}
  \country{Austria}
  \postcode{1090}
}

\author{Torsten Möller}
\email{torsten.moeller@univie.ac.at}
\affiliation{%
  \institution{University of Vienna, Faculty of Computer Science, Research Group Visualization and Data Analysis, Research Network Data Science}
  \streetaddress{Sensengasse 6}
  \city{Vienna}
  \state{Vienna}
  \country{Austria}
  \postcode{1090}
}

\author{Sebastian Tschiatschek}
\email{sebastian.tschiatschek@univie.ac.at}
\affiliation{%
  \institution{University of Vienna, Faculty of Computer Science, Research Network Data Science}
  \streetaddress{Währinger Straße 29}
  \city{Vienna}
  \state{Vienna}
  \country{Austria}
  \postcode{1090}
}

\renewcommand{\shortauthors}{Schmude et al.}

\begin{abstract}
      Every AI system that makes decisions about people has a group of stakeholders that are personally affected by these decisions. However, explanations of AI systems rarely address the information needs of this stakeholder group, who often are AI novices. This creates a gap between conveyed information and information that matters to those who are impacted by the system's decisions, such as domain experts and decision subjects. To address this, we present the "XAI Novice Question Bank," an extension of the XAI Question Bank~\cite{liao2020} containing a catalog of information needs from AI novices in two use cases: employment prediction and health monitoring. The catalog covers the categories of data, system context, system usage, and system specifications. We gathered information needs through task-based interviews where participants asked questions about two AI systems to decide on their adoption and received verbal explanations in response. Our analysis showed that participants' confidence increased after receiving explanations but that their understanding faced challenges. These included difficulties in locating information and in assessing their own understanding, as well as attempts to outsource understanding. Additionally, participants' prior perceptions of the systems' risks and benefits influenced their information needs. Participants who perceived high risks sought explanations about the intentions behind a system's deployment, while those who perceived low risks rather asked about the system's operation. Our work aims to support the inclusion of AI novices in explainability efforts by highlighting their information needs, aims, and challenges. We summarize our findings as five key implications that can inform the design of future explanations for lay stakeholder audiences. 
\end{abstract}




\maketitle

\def\thesection{\arabic{section}}

\section{Motivation}
\label{sec:motivation}



Society is in the process of negotiating which AI systems can be used responsibly in public institutions and organizations~\cite{zuger_ai_2023, lucaj2023, lima2023}. These systems should be transparent, accountable, and support human oversight, according to the legal frameworks that are being developed by regulatory bodies~\cite{european_commission_laying_2021}.  Research in the field of explainable AI (XAI) addresses the realization of these values by examining how AI systems can be made understandable to stakeholders and how the given information enables specific actions, such as the contestation of decisions~\cite{Alfrink2022} and the assignment of accountability~\cite{langer_what_2021}. As explanations thereby assume a key role in the deployment of responsible AI systems for the public~\cite{zuger_ai_2023}, the question of what constitutes a \textit{good} explanation becomes central.  

Good explanations of AI systems are aligned with the information needs of the stakeholders~\cite{miller_explanation_2019, mueller2019explanation, byrne_good_2023, ehsan_human-centered_2022} and are a means to help them achieve their aims, which can depend on the stakeholder's role~\cite{langer_what_2021, arrietta2020}, AI literacy~\cite{long_what_2020}, and domain knowledge~\cite{wang21}. However, information needs of people who are impacted directly and personally by algorithmic decision-making (ADM) systems~\cite{european_parliament_understanding_ADM_2019, golpayegani_be_2023}, such as domain expert users and decision subjects, have not yet been explored in-depth in XAI research. As a result, current explanations rarely meet the needs of these stakeholder groups. This is critical, as providing explanations for these stakeholder groups can be essential to reduce the risks of power asymmetry~\cite{ananny_seeing_2018}, inequality~\cite{lopez_reinforcing_2019}, and informational unfairness~\cite{schoeffer_there_2022} in deployments of public AI~\cite{kuziemski2020, henman_improving_2020, brown_toward_2019}.

However, building explanations that aid domain experts and decision subjects in understanding ADM systems can be challenging. First, these stakeholder groups are diverse and can include "AI novices"\footnote{We use the term in reference to "visualization novices"~\cite{burns2023} and synonymously with "lay people"~\cite{shen_designing_2020, szymanski_visual_2021, schmude2023}.}, i.e., people with little technical literacy who have heterogeneous information needs and prior knowledge, which necessitates accessible and adaptive explanations~\cite{shulner-tal_enhancing_2022, conati_toward_2021, KIM2024103160}. Second, the effects of explanations are influenced by perceptions of the deploying institution~\cite{brown_toward_2019, ehsan_expanding_2021}, meaning that if the institution is not trusted, the ADM system and its explanations are not trusted either~\cite{woodruff_qualitative_2018, scott_algorithmic_2022}. %
Third, understanding is an interconnection of knowledge and reasoning~\cite{grimm_varieties_2019, keil2006} involving numerous cognitive processes~\cite{keil2006, lombrozo_mechanistic_2019} that can barely be captured by standardized large-scale tests~\cite{sato_testing_2019}. Lastly, AI lifecycles encompass large amounts of information~\cite{dhanorkar_who_2021} and, therefore, require a preselection of what is deemed relevant.

These interactions between cognitive, perceptional, and organizational dimensions can be difficult to address in explanation design. 
Recent XAI research, therefore, advocates for providing tailored explanations, based on the stakeholder's aims~\cite{dhanorkar_who_2021, freiesleben2023} and information needs~\cite{liao2020, liao2021questiondriven}, instead of explaining what is assumed to be relevant~\cite{miller2023}. This study thus follows research that aims to build "human-centered" explanations of AI by making them more accessible and better suited to human cognitive processes~\cite{shin_algorithms_2023, ehsan_human-centered_2022}.   

This paper contributes to including domain experts and decision subjects in explainability by presenting an empirical study of their information needs, understanding, and perceptions. We use two ADM use cases as examples: an employment prediction algorithm and a health wristband for geriatric care. To collect participants' information needs, the explanation setting is flipped: Participants ask questions about the algorithmic systems to evaluate their adoption and are given verbal explanations in response. Participants thus independently choose which information they inquire about, reducing the amount of pre-selected information in favor of demand-based information. We analyze the impact of question-driven explanations on participants' self-reported understanding and decision confidence, as well as the role of "explanatory stances"~\cite{keil2006} in their information acquisition. Lastly, we analyze participants' perceptions of risks and benefits before and after receiving explanations and describe perceived changes. Our work is guided by the following research questions: 

\begin{enumerate}[label={[RQ\arabic*]}, leftmargin=10ex]
    \item \emph{Information Needs}: What information do AI novices who could be affected by algorithmic decisions need in order to decide about adopting an ADM system? 
    \item \emph{Understanding}: How do question-driven explanations interact with participants' understanding? 
    \item \emph{Perception}: How do question-driven explanations impact perceptions of the systems' risks and benefits? 
\end{enumerate}

The contributions of our work include i) a catalog of affected stakeholders' information needs summarized in the "XAI Novice Question Bank" (\autoref{fig:xainqb})\footnote{In reference to~\citet{liao2020}'s XAI Question Bank.}; ii) recommendations on how to address understanding challenges such as blind spots and outsourcing and on how to use explanatory stances in question-driven explanations; and iii) reflections on the relevance of information about intention in explanations for affected stakeholders and considerations of how to incorporate insights from "civic education"~\cite{lupia_uninformed_2016} into explainability.

\section{Background and related work}
\label{sec:related_work}

We structured our literature review in three sections. The first section describes related work in Human-Centered AI on high-risk systems and stakeholder-focused explainability (Section~\ref{sec:hcai}). The second section covers the interplay between explanations and understanding as well as definitions of understanding and approaches to analyze it (Section~\ref{sec:explanations-and-understanding}). The third section describes the importance of people's perceptions of ADM systems' risks and benefits and the relationship between explanations and the assessment of normative values (Section~\ref{sec:perceptions-of-risks}).

\subsection{Human-Centered AI (HCAI)}
\label{sec:hcai}

The notion of human-centered AI aims to provide principles and frameworks that enable "ethical, interactive, and contestable use" of AI systems, covering both interaction design approaches as well as the incorporation of ethical values~\cite{capel_what_2023}. \citet{ShneidermanBen2022HA}, in line with other scholars~\cite{Araujo2020, xu2019, shin_algorithms_2023, shin2023_hcai}, emphasizes that human-centered AI systems should 1) build on user observation and stakeholder engagement and should 2) empower rather than replace people, i.e., be "accessible and controllable"~\citet{shin2023_hcai}. Prioritizing people, in this sense, means giving attention to users and other stakeholders throughout the development, deployment, and evaluation of AI systems~\cite{shin_algorithms_2023} while observing human values such as fairness, trust, and accountability~\cite{ShneidermanBen2022HA}. Similar notions have been brought forward in multiple responsible and trustworthy AI guidelines, such as the Montreal Declaration for Responsible Development of AI, the Vienna Manifesto on Digital Humanism, and, more recently, the EU AI Act~\cite{european_commission_laying_2021}. 

Human-centered design becomes especially important when considering "high-risk" AI applications -- systems that pose risks to the health, safety, or fundamental rights of persons~\cite{european_commission_laying_2021}. Rendering high-risk systems transparent, understandable, and explainable to users and preserving their control is seen as a mitigation of these risks~\cite{adadi_peeking_2018}, but the practical realization proves challenging~\cite{ananny_seeing_2018}. As people who are affected by ADM systems tend to be AI novices, explanation approaches need to be tailored to their needs, but as~\citet{shin_algorithms_2023} notes, there are few "well-established ways of incorporating lay users into the process of designing algorithms". Further, no universal explanation method exists that can cover the needs of all stakeholder groups equally~\cite{phillips_four_2021}. Therefore, the question of how AI novices' information needs can be analyzed and incorporated into explanation approaches is an open challenge in XAI research~\cite{shulner-tal_enhancing_2022}. 

To this end, several studies have examined how explanations can help lay users to understand algorithmic systems: \citet{szymanski_visual_2021} examined explanations for model predictions and found that lay users preferred a visual format over a textual one but had difficulties correctly interpreting the visual explanation, which led to the development of a new explanation combining both formats. \citet{cheng_explaining_2019} found that interactive explanations and white-box models improved lay participants' comprehension; however, a later study by \citet{bove2022} could not replicate the results regarding interactive explanations and found the adverse effect. \citet{bove2022} suggest that the diverging results could have stemmed from the difference in task domains (student admission vs.\ car insurance) and emphasize that context influences the explanation's effect. Lastly, \citet{bertrand2023} used interviews to elicit the information needs of domain experts and end-users about a financial robo-advisor, which they then used to develop and test explanations for the system's recommendations in a large-scale survey. Their analysis showed that feature-based SHAP explanations, one of the most well-known explanation methods, did not increase users' understanding nor enable a suitable trust calibration. In contrast, our study takes a more general perspective on the qualitative exploration of participants' global information needs and perceptions when they vote on the adoption of a high-risk ADM system.   

These multi-faceted insights indicate how difficult it can be to tailor explanations of algorithmic systems to lay people. For this reason, this study aims to take a closer look at the information needs of AI novices and how they can be categorized. To elicit these needs, participants in our study assume the role of a decision-maker who can ask questions and receive explanations in return, as described in Section~\ref{sec:method}. We thus "flip" the explanation setting to answer the following research question:  

\textit{RQ1-Information Needs: What information do AI novices who could be affected by algorithmic decisions need in order to decide about adopting an ADM system?}

\subsection{Explanations and understanding}
\label{sec:explanations-and-understanding}

One purpose of explanations is to increase the recipient's \textit{understanding} of a system~\cite{langer_what_2021}. This increased understanding can help people to pursue their own goals, such as to assess the system's values~\cite{kim2023}, consider its recommendations~\cite{lee_webuildai_2019}, or contest its decisions~\cite{Alfrink2022}. \citet{shin_algorithm_2022} further state that the understandability of explanations impacts how users evaluate both the quality of explanations and the system's fairness, accountability, and transparency. However, despite this vital role of understanding, there is no general conception of what is most important about an AI system, nor how this understanding can be evaluated~\cite{schmude2023}.   

Numerous definitions of understanding have been proposed in the literature~\cite{grimm_varieties_2019, baumberger_what_2017, zagzebski_toward_2019, keil2006}. For this work, we use a definition that aims to preserve its epistemic aspects while allowing for operationalization in an empirical setting. We thus define understanding as i) connecting and applying information, whereas "knowledge" only stores information~\cite{grimm_varieties_2019, baumberger_what_2017}, ii) involving different varieties depending on whether people (intention, norms) or objects (causality, natural laws) are understood~\cite{grimm_varieties_2019}, and iii) being the attempt to grasp the underlying structure of a phenomenon by way of simplification~\cite{zagzebski_toward_2019}.  

\subsubsection{Gaps in understanding and explanatory stances}
To integrate the epistemic aspects of understanding practically, we use models provided by the cognitive sciences.~\citet{keil2006} and~\citet{lombrozo_mechanistic_2019} state that explanations increase understanding by calling attention to \textit{gaps in understanding}. These gaps can be dealt with by (i) filling them with new information, (ii) "outsourcing" them to another mind, or (iii) deciding that they are irrelevant. In this sense, developing understanding means dealing with understanding gaps until one feels to have reached a "working understanding"~\cite{keil_folkscience_2003, keil2006}. This process can be perturbed by the "illusion of explanatory depth"~\cite{rozenblit2002}, an overestimation of one's understanding, which can lead to incomplete understanding~\cite{keil2006, lombrozo_mechanistic_2019, chromik_i_2021}.

For our analysis, we further use the concept of three \textit{explanatory stances} \cite{keil_2021_mechanistic_explanation, keil2006, lombrozo_explanation_2014, lombrozo_mechanistic_2019}, introduced by~\citet{dennett_intentional_1998}: a theory aiming to capture how people make sense of information and generalize from it in order to develop predictive strategies. The three explanatory stances are described below in the order \textit{mechanical}, \textit{design}, and \textit{intentional} stance. 

\begin{itemize}
    \item The \textit{mechanical stance} focuses on low-level technical detail, such as "parts, processes, and proximate causal mechanisms"~\cite{lombrozo_mechanistic_2019}. For example, explaining why an alarm clock buzzes in a mechanical stance would state that the circuit connecting the power source and buzzer was completed~\cite{lombrozo_mechanistic_2019}. In contrast to functional and intentional explanations, mechanistic\footnote{Both the terms ’mechanical’ and ’mechanistic’ are used in the literature~\cite{keil2006, lombrozo_mechanistic_2019, keil_2021_mechanistic_explanation}. Where ’mechanical’ refers to the stance, ’mechanistic’ refers to the explanation.} explanations do not take into account aims or beliefs~\cite{Paez2019}.
    \item The \textit{design stance} focuses on functions and goals beyond mechanical interactions in a system~\cite{keil2006} and describes why things happen by looking at their functional purposes~\cite{lombrozo_explanation_2014}. Explaining why an alarm clock buzzes from a design stance would state that it functions according to its design: waking its owner at the set time~\cite{lombrozo_mechanistic_2019, Dennett2006-DENIST}.
    \item The \textit{intentional stance} focuses on a system's purpose as defined by the human beliefs or aims ingrained in it~\cite{keil2006, dennett_intentional_1998, yurrita2022}. Explaining why an alarm clock buzzes in an intentional stance might refer to its purpose of waking its owner, whose desire is to have enough time before leaving for work. 
\end{itemize}

Previous XAI literature used explanatory stances to examine how explanations are perceived and evaluated~\cite{Paez2019, zerilli_2022, miller_explanation_2019, byrne_good_2023}. For example, \citet{byrne_good_2023} describes that explanatory stances shed light on how people explain the behavior of algorithmic systems to themselves, which might differ depending on whether they interpret the decisions as results of an algorithmic process or as depictions of decisions by a human (e.g., a job counselor). The stances, therefore, aid in identifying whether the users' information needs focus on the systems' technical or intentional aspects and thus make it easier to match their information needs to a specific explanation~\cite{byrne_good_2023}. In our approach, we use the stances to differentiate if participants asked for information about technical details, functional relations, or the sociotechnical context.
  
\subsubsection{Question-driven explanations}

Questions assume a central role in XAI, as explanations of AI systems can be defined as responses to questions that are asked by the user.~\citet{miller_explanation_2019} defines three classes of these questions (what, how, and why), which are meant to cover different levels of causal reasoning through their responses. A comparable classification approach has been developed earlier by~\citet{lim+dey_assessing_intelligibility2009}, who defined five intelligibility questions (what, why, why not, what if, and how to) to capture questions that "end-users of novel systems may ask" to convey to users the practical functionality of these systems. A practical implementation of question-driven explanation was pursued even earlier by~\citet{ram1989}, who built AQUA -- a system that used questions about a story's content to generate explanations and advance its knowledge.

From a didactic perspective,~\citet{wiggins_understanding_2005} describe questions to be "doorways to understanding", through which learners explore concepts, theories, and problems that reside within the content. When we thus define explanations as answers to questions, it is noteworthy that these questions often remain tacit in XAI. In response,~\citet{liao2021questiondriven} propose a question-driven design process for explanations, which is grounded in the needs and questions of users. The concept further builds upon their previous work on the XAI Question Bank~\cite{liao2020}, a collection of expert users' information needs represented as prototypical questions (e.g., "What kind of output does the system give?"). In our setting, we apply the question-driven design process by providing participants with explanations as responses to their questions. We aim to examine how participants' understanding is changed by these explanations and whether they allow for a better matching between information and explanatory stances. To this end, we pose the following research question: 

\textit{RQ2-Understanding: How do question-driven explanations interact with participants' understanding? }

\subsection{Perceptions of risks and benefits of algorithmic systems}
\label{sec:perceptions-of-risks}

The acceptance and responsible use of algorithmic systems in society majorly depends on how these systems are perceived~\cite{jakesch_how_2022, shin_credibility_2022, zuger_ai_2023}. If the "technical affordances" of a system diverge from the "social needs"~\cite{ehsan2023}, i.e., if the system does not or cannot do what is expected from it, it can result in users' distrust in the deploying institution and their rejection of the system~\cite{jakesch_how_2022}. Explainability is understood to influence these perceptions, taking a "facilitating" role in how users evaluate an algorithm's features and their attitudes towards it~\cite{SHIN2021, langer_what_2021}. \citet{shin_algorithms_2023} describes this as a "dual-step flow", in which explainability allows users to determine their perceptions of the system's fairness, transparency, and accountability (non-functional attributes), which then, in turn, lets them calibrate their trust in the system and their evaluation of accuracy and personalization (functional attributes). Trust, in this sense, takes a mediating role in that it connects explainability to the system's evaluation by the user~\cite{shin_credibility_2022}.

Explanations thus play an essential role in helping people adjust their perceptions of whether an algorithmic system adheres to normative values. But to fulfill this facilitating function, the information provided in explanations needs to match "what users really wish to understand"~\cite{shin_algorithms_2023}. Ensuring this fit can further improve the perceived "informational fairness" of algorithmic systems, meaning that information is consistent, reveals relationships between input and output, and includes actionable insights~\cite{schoeffer_there_2022}. However, several studies have shown the complex interrelations that exist between normative values, trust, acceptance, and explanations~\cite{shulner-tal_enhancing_2022, shin_understanding_2022, schoeffer_there_2022}, and the meaning of normative values such as fairness, trust, and transparency is not strictly fixated in colloquial speech, making it difficult to compare self-reports by participants~\cite{schmude2023}. Motivated by~\citet{corvite_data_2023}'s work, we therefore use an evaluation based on perceived risks and benefits of algorithmic systems. By using a two-dimensional scale that depicts both risk and benefit simultaneously (described in Section~\ref{sec:method}), we further aim to let participants consider the positive and negative consequences of deploying a specific algorithmic system. As we focus on the explanations' effect on these perceptions, we formulate the third research question: 

\textit{RQ3-Perception: How do question-driven explanations impact perceptions of the systems' risks and benefits?}

\section{Method}
\label{sec:method}

To elicit the information needs of AI novices, we conducted an interview study in German with 24 participants recruited from a local job agency and a local apartment complex inhabited mainly by retirees (Section~\ref{sec:participants}). We used a qualitative approach for this study to allow for comprehensive conversations with participants and the collection of data that could be analyzed thematically~\cite{braun_using_2006}. The interviews were thus designed to develop a theoretical account of information needs that is grounded in empirical observation~\cite{urquhart_putting_2009} (Section~\ref{sec:analysis}). The focus of this study is therefore on the conceptualization and exploration of AI novices' information needs, which aims to lay the groundwork for further qualitative and quantitative examination in future work. 

We presented participants with one of two algorithmic decision-making (ADM) use cases (Section~\ref{sec:use_cases}): an employment prediction algorithm or a health wristband for geriatric care. Participants were tasked to decide whether the presented use cases should be adopted. Before deciding, participants had the opportunity to gather information about the systems by asking questions in two 15-minute inquiry phases. In response to their questions, they received verbal explanations from the main author of this paper. The study closed with an interview at around 50 minutes. The data was analyzed using inductive and deductive analysis (as described in Section~\ref{sec:analysis}). 

The procedure was designed to fulfill multiple study goals. Firstly, the decision task was meant to give participants a sense of importance for their decision and to motivate them to acquire relevant information. Secondly, by letting participants inquire freely and flexibly about the use cases in the first phase and providing the XAI Question Bank~\cite{liao2020} only in the second phase, we could examine both participants' intuitive information needs and the bank's impact on their inquiry. Third, by eliciting participants' understanding of the ADM systems and their perceptions of the risks and benefits, we can examine the interactions between these aspects and participants' information needs. The whole study setup is described in Section~\ref{sec:Study_setup}, and the study procedure is depicted in~\autoref{fig:procedure}. 

\subsection{Use cases}
\label{sec:use_cases}

The two use cases were selected as representative forms of so-called "high-risk" ADM systems, i.e., systems that pose risks to the health, safety, or fundamental rights of people~\cite{european_commission_laying_2021}. In particular, they are examples of the domains \textit{employment} and \textit{biometrics}, which are defined as areas with high-risk potential in the upcoming EU AI Act~\cite{european_commission_laying_2021}.  

Both use cases are thus examples of more general application areas of high-risk algorithmic systems. While models, data, and context are specific to each use case, the overarching themes of information inquiry and the utility of question-driven explanations are broadly applicable. They can be transferred to the explanation design for other algorithmic systems. In the following, each use case is briefly described, and its selection is motivated. 

\subsubsection{Use Case A: AMS (employment prediction) algorithm} 

\textit{Description.} The \amsalgorithm\footnote{AMS stands for the Public Employment Agency (Arbeitsmarktservice).} is a scoring system meant to predict the employability of job-seekers in Austria. It was developed by a private company for the Austrian Public Employment Agency between 2015 and 2021 but was never used as a live system and was put on hold in 2021 due to privacy objections~\cite{allhutter_bericht_ams-algorithmus_2020}. To predict employability chances, the system used a logistic regression model trained on historical data to calculate how job-seekers' demographic features influenced their chances (such as age, education, nationality, etc.\footnote{The full list of features is given in the supplementary material.}). This model would produce a short-term and long-term employment score for each job-seeker. The scores would be given to the job-seeker's counselor at the employment agency to aid in deciding about suitable support measures. Counselors could overwrite the system's group assignment of job-seekers but would need to give a reason for doing so~\cite{allhutter_bericht_ams-algorithmus_2020, Holl2018}. 

\textit{Choice of use case.} 
Algorithmic tools for the profiling of job-seekers have been deployed in various countries, including Germany~\cite{agentur_für_arbeit_2021}, Austria~\cite{allhutter_bericht_ams-algorithmus_2020}, Poland~\cite{niklas2015}, and the Netherlands~\cite{DESIERE_STRUYVEN_2021}. The~\amsalgorithm is thus an example of a more general trend: the introduction of automation and algorithmic procedures into public employment agencies to increase productivity and efficiency~\cite{Allhutter2020en}. Profiling tools like the~\amsalgorithm assist in assessing job-seekers and resource allocation and often split the responsibility for tasks between the algorithmic system and caseworkers~\cite{scott_algorithmic_2022}. 
However, the introduction of these applications also repeatedly highlights which parts of human interaction cannot be easily replaced by automation, such as "the desire to be seen as a whole human being"~\cite{scott_algorithmic_2022}. As the \amsalgorithm, therefore, conjoins typical procedures, interactions, and challenges of ADM systems in public employment, it presented a suitable use case for our study.

\subsubsection{Use Case B: Health wristband for geriatric care} 

\textit{Description.}
The health wristband is an assistance tool for the long-term care of older adults, which tracks and evaluates movement and acceleration data and raises an alarm when the wearer trips or falls. The system does not require manual activation by the wearer but relies on trained machine learning models to detect events (erratic movement, heart rate anomalies, etc.) in the recorded data, making it viable for use in private homes. Further, the wristband collects data on the wearer's sleeping, movement, and eating behavior, allowing for remote monitoring of patients. The wristband thus fulfills a two-part purpose for the care organization: First, harmful events can be detected immediately, and thus, little time is lost when dispatching emergency services. Second, patients' data can be used for the organization of care services by making overall coordination more efficient. Systems with comparable functionality have reportedly been employed in Sweden since 2016~\cite{kaun2020, sahlen2018}.

\textit{Choice of use case.} 
Wearable devices for identifying and monitoring patients are already in use in many care organizations~\cite{kutsarova2020, kaun2020, sahlen2018}, such as nursing homes~\cite{chang2023} and psychiatric wards~\cite{NahavandiDarius2022Aoai}. Wristbands, in particular, can capture signals such as heart rate and body temperature while tracking movement and sleep~\cite{seneviratne2017}. This data can then be used to detect anomalies in breathing, physical movement, or other vital signs and trigger the appropriate reaction. This facilitates health status monitoring without keeping patients in the hospital or nursing homes, making them useful for various scenarios~\cite{kutsarova2020}. Besides being an example of wearable devices already well-established in care contexts, the wristband also incorporates a direct trade-off between perceived privacy and safety, as it can help detect risky situations in exchange for continuous monitoring~\cite{kutsarova2020}. Consequently, some users might reject these health devices due to feeling under surveillance~\cite{NahavandiDarius2022Aoai}. A further known issue is prediction inaccuracy for certain population groups due to biases in the training data~\cite{GERKE2020295}. As the wristband is thus representative of known benefits and risks of wearable devices in care contexts, we selected it as the second use case for our study. 

\subsection{Study procedure and setup}
\label{sec:Study_setup}

\begin{figure*}[h]
    \centering
    \includegraphics[width=\textwidth]{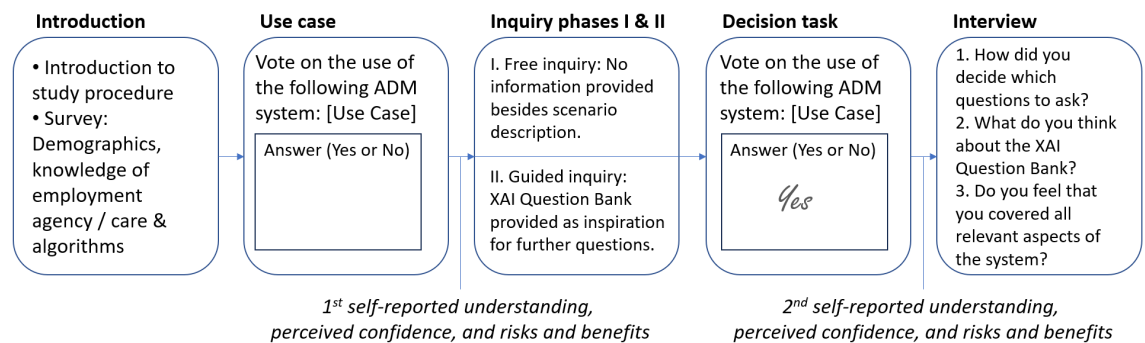}
    \caption[Study procedure]{\textbf{Depiction of the study procedure.} Participants received a short introduction to the study, filled out a questionnaire on demographic information and prior knowledge, and then were presented with one of two use cases. Participants could then ask questions about the given use case in two 15-minute inquiry phases. In the second of these phases, they received the XAI Question Bank~\cite{liao2020} for reference. They then voted on adopting the discussed system and answered three interview questions. Participants were asked two times for their self-reported understanding, perceived decision confidence, and perceived risks and benefits of the system: before the inquiry phases and after making the decision.}
    \label{fig:procedure}
\end{figure*}

The study procedure consisted of several steps that are depicted in \autoref{fig:procedure}. This section describes the use cases and decision task, the inquiry phases I and II, the question-driven explanations during the study and the reporting mechanisms for understanding, confidence, and risks and benefits.

\subsubsection{Use case and decision task}

At the beginning of the study, participants were informed that they would have to vote over the adoption of the ADM system after the inquiry phases, which introduced an incentive to gain information. After participants were given a consent form and a questionnaire on demographical information and previous knowledge, they were presented with the decision use case, including a brief description of the use case and its planned deployment. The description rendered the participants as part of a small citizen referendum asked to provide their vote on the system. A small box was printed below the use case description for participants to fill in "yes" or "no" after the inquiry phases, acting like a voting ballot. The design was inspired by ballots used in popular votes in Switzerland~\cite{switzerland_voting}.\footnote{Popular votes in Switzerland require the voter to write "Yes" or "No" on the ballot, which can be argued to be more deliberate compared to a tick box.}     

\subsubsection{Inquiry phases I \& II}

After being briefed about the decision use case, participants had 30 minutes to ask the study examiner questions and receive verbal explanations about the ADM use case before making their decision. The time available to participants for inquiries about the use cases was split into two phases of 15 minutes. In the first phase, participants asked questions about the use case from their own intuition. In the second phase, participants were provided with a reduced version of the XAI Question Bank~\cite{liao2020}\footnote{A detailed tabular explanation for the reduction is given in the supplementary material.}, a collection of questions on AI systems sourced from AI developers and designers, which they could use as a reference for further questions. As the original XAI Question Bank includes a comprehensive collection of questions, we reduced it by summarizing questions that aim toward similar topics. Participants were given a pen and paper to take notes during their inquiry.  

\subsubsection{Question-driven verbal explanations}

\begin{figure}[]
    \centering
    \includegraphics[width=\textwidth]{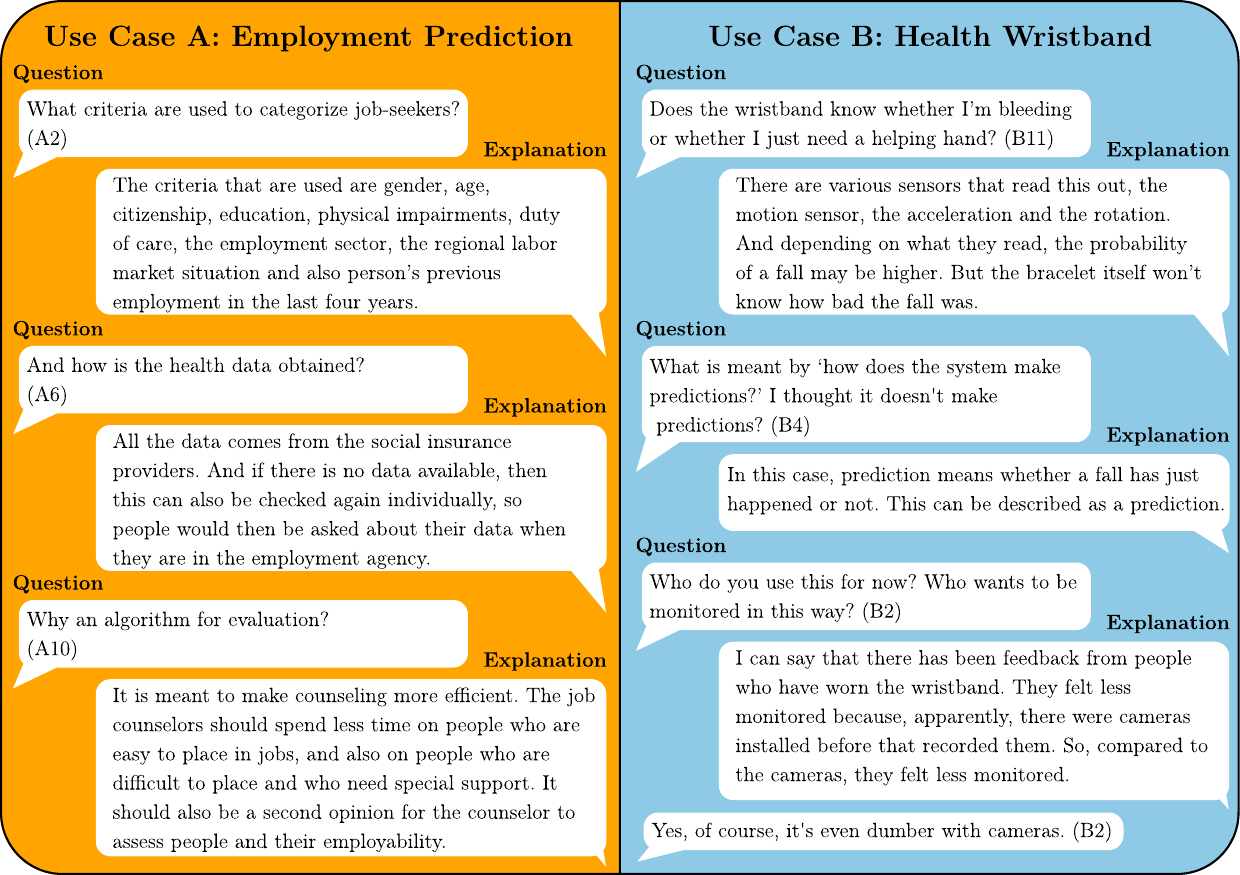}
    \caption[Questions and answers]{\textbf{Interview transcripts showing examples of question-driven explanations.} The study examiner responded to participants' questions with verbal explanations during the inquiry phases I and II. All explanations were based on publicly  available information about the systems and were phrased so as not to convey any personal opinion or judgment.}
    \label{fig:q&a}
\end{figure}

Participants received verbal explanations from the study examiner as responses to their questions. The explanations were based on information prepared in advance and available to the study examiner during the study (as depicted in~\autoref{fig:q&a}). For each use case, two pilot studies were conducted to collect a stock of questions that would likely be asked in the subsequent studies and which served as guidance for preparing information. When questions could not be answered due to a lack of information, the examiner asked how and why the requested information would be relevant to participants (for examples, please refer to the supplementary material).

\subsubsection{Self-reported understanding, confidence, and risks \& benefits}
\label{sec:method_self_reports}

Participants reported their understanding, decision confidence, and perceptions of the system's risks and benefits (and an explanation for their answer) at two points: first, before the inquiry phases, and second, after making the decision. Self-reported understanding and decision confidence were elicited through 5-point Likert scales. To encourage participants to speak about the trade-offs between the systems' risks and benefits, these were split into societal and personal dimensions. For each elicitation of risks and benefits, two visual analog scales (10 cm)~\cite{rugg2007} were combined to form a two-dimensional Cartesian coordinate system with the origin located at 5 cm on both axes. By denoting perceived risk on the x-axis and perceived benefit on the y-axis, the coordinate system formed four quadrants of varying risk and benefits combinations (as depicted in~\autoref{fig:r&b_society}). Participants drew marks on the axes, measured, and then rounded to the first decimal place. 

\subsection{Interview method and analysis}
\label{sec:analysis}

\textit{Interview method.} The study employed two interview techniques throughout the study procedure (\autoref{fig:procedure}). In inquiry phases I and II, participants were free to ask any question about the ADM use case that they found relevant. This led to a form of \textit{unstructured interview}, in which not the examiner but the participant was driving information acquisition. At this point, the examiner only asked questions to clarify participants' inquiries and to determine why they asked for certain information. In the last part of the study, the examiner conducted a \textit{semi-structured interview} to gather participants' responses on three key aspects: their overall reasoning and strategy for asking questions, their perception of the XAI Question Bank's~\cite{liao2020} usefulness, and their perception of whether they acquired a complete picture of the system. At this point, the study examiner followed up on thematic aspects that arose during the interview and further connected participant responses to the quantitative items elicited before to facilitate qualitative exploration~\cite{Weiss1995}. 

\textit{Analysis.} Regarding RQ1-Information Needs, we use thematic analysis~\cite{braun_using_2006} to examine participants' information needs inductively and deductively using the XAI Question Bank~\cite{liao2020}. All questions raised by participants in the two inquiry phases were coded according to the category of inquiry. Categories already present in the XAI Question Bank~\cite{liao2020} were assigned deductively, and topics not yet covered were defined inductively from the data. The XAI Question Bank~\cite{liao2020} thus served as a starting point for a comparative analysis between participants' information needs and those of AI practitioners, highlighting similarities and discrepancies between the groups. A team of three coders from the authors' research group checked an intermediary version of the coding for inter-coder agreement by independently analyzing the transcripts and comparing code categories.

For RQ2-Understanding, we evaluated participants' responses on the self-reported understanding and decision confidence Likert scales and combined this with a deductive analysis using the frameworks of "explanatory stances"\cite{keil2006} and understanding challenges~\cite{wiggins_understanding_2005, keil2006} (described in Section~\ref{sec:related_work}). 

For RQ3-Perception, we analyze how explanations provided during the inquiry phases impact participants' perceptions of the systems' risks and benefits by comparing their 
reports on the risks and benefits coordinate system (Section \ref{sec:method_self_reports}) before and after inquiry and analyzing their corresponding articulations. Further, we analyze how participants' perceptions influence which questions are asked by comparing their reports on risks and benefits and information needs between use cases.  

\subsection{Participants}
\label{sec:participants}

\textit{Recruitment.}~\autoref{fig:Participant_table} provides an overview of the study participants. We recruited participants from two main locations: In the employment prediction use case, we collaborated with the local employment agency Job-TransFair, which established contact with job-seekers and personnel counselors. In the wristband use case, we used street sampling in an apartment complex known for the quality of living it offers to retirees.\footnote{While we contacted many institutions active in geriatric care work, none wanted to collaborate. The reasons for this are unknown. However, the difficulty of recruiting retirees for study participation over official channels is noteworthy.} The recruitment criteria were i) no or little previous experience with either algorithms or AI systems (reported in the questionnaire) and ii) identification with either the domain expert or decision subject stakeholder group as described below. All interviews were conducted in person, either in office spaces or cafés, except for one conducted online. Participants were compensated with \euro{}~12 in cash, which was disclosed only after completion of the study so as not to incentivize participation solely for compensation.

\textit{Choice of participants.} In both use cases, participants were selected such that they were representative for one of two roles: \textbf{domain experts} or \textbf{decision subjects}. Following the XAI stakeholder framework~\cite{arrietta2020, langer_what_2021}, we define domain experts as people who are competent in the field that the ADM system is used in (e.g., employment and healthcare) and who would be its likely users. As domain experts, we recruited personnel counselors for the employment prediction use case, and care workers and a doctor for the health wristband use case. All domain experts had multiple years of experience in their field and could readily imagine how the corresponding ADM system would integrate with their work.

We describe decision subjects as people who are impacted by an automated decision and might seek ways to contest or change it but who are not necessarily users of the system~\cite{Alfrink2022, yurrita2022}. We recruited job-seekers and people who had previously been job-seeking as decision subjects for the employment prediction use case, as well as retired people and their relatives for the health wristband use case. Most of these participants already had experience with systems comparable to our use cases, such as system providing statistical competence analyses of job-seekers and wearable alarm systems for retired people in geriatric care, which, however, did not include machine learning.   

\begin{table}[H]
\centering
\caption{Details on the study participants. The ID indicates which use case participants were presented with: A = \textit{employment prediction}, B = \textit{health wristband}. The decision column shows how participants voted regarding the adoption of the corresponding ADM system.}
\resizebox{\textwidth}{!}{%
\begin{tabular}{lllll|llllll}
\hline
ID  & Age & Education             & Occupation           & Decision &  & ID  & Age & Education             & Occupation     & Decision \\ \hline
A1  & 54  & University            & Personnel counselor & Yes      &  & B1  & 62  & A level               & Retired        & Yes      \\
A2  & 61  & Vocational university & Personnel counselor & No       &  & B2  & 68  & A level               & Retired        & Yes      \\
A3  & 54  & Secondary school      & Job-seeking          & Yes      &  & B3  & 57  & Secondary school      & Retired        & Yes      \\
A4  & 43  & A level               & Personnel counselor & Yes      &  & B4  & 46  & Apprenticeship        & Geriatric care & No       \\
A5  & 27  & Apprenticeship        & Job-seeking          & No       &  & B5  & 80  & Secondary school      & Retired        & Yes      \\
A6  & 53  & University            & Personnel counselor & No       &  & B6  & 68  & University            & Retired        & Yes      \\
A7  & 63  & University            & Personnel counselor & n/a      &  & B7  & 58  & University            & Freelancer     & Yes      \\
A8  & 56  & Apprenticeship        & Job-seeking          & No       &  & B8  & 70  & Vocational university & Retired        & Yes      \\
A9  & 50  & A level               & Job-seeking          & No       &  & B9  & 36  & University            & Geriatric care & Yes      \\
A10 & 63  & Secondary school      & Cleaning service     & No       &  & B10 & 66  & Secondary school      & Retired        & Yes      \\
A11 & 40  & A level               & Personnel counselor & Yes      &  & B11 & 89  & Apprenticeship        & Retired        & Yes      \\
A12 & 25  & Secondary school      & Job-seeking          & Yes      &  & B12 & 60  & University            & Doctor         & Yes     \\ \hline
\end{tabular}}
\label{fig:Participant_table}
\end{table}

\section{Results}
\label{sec:results}

\begin{figure}[]
    \centering
    \includegraphics[width=400pt]{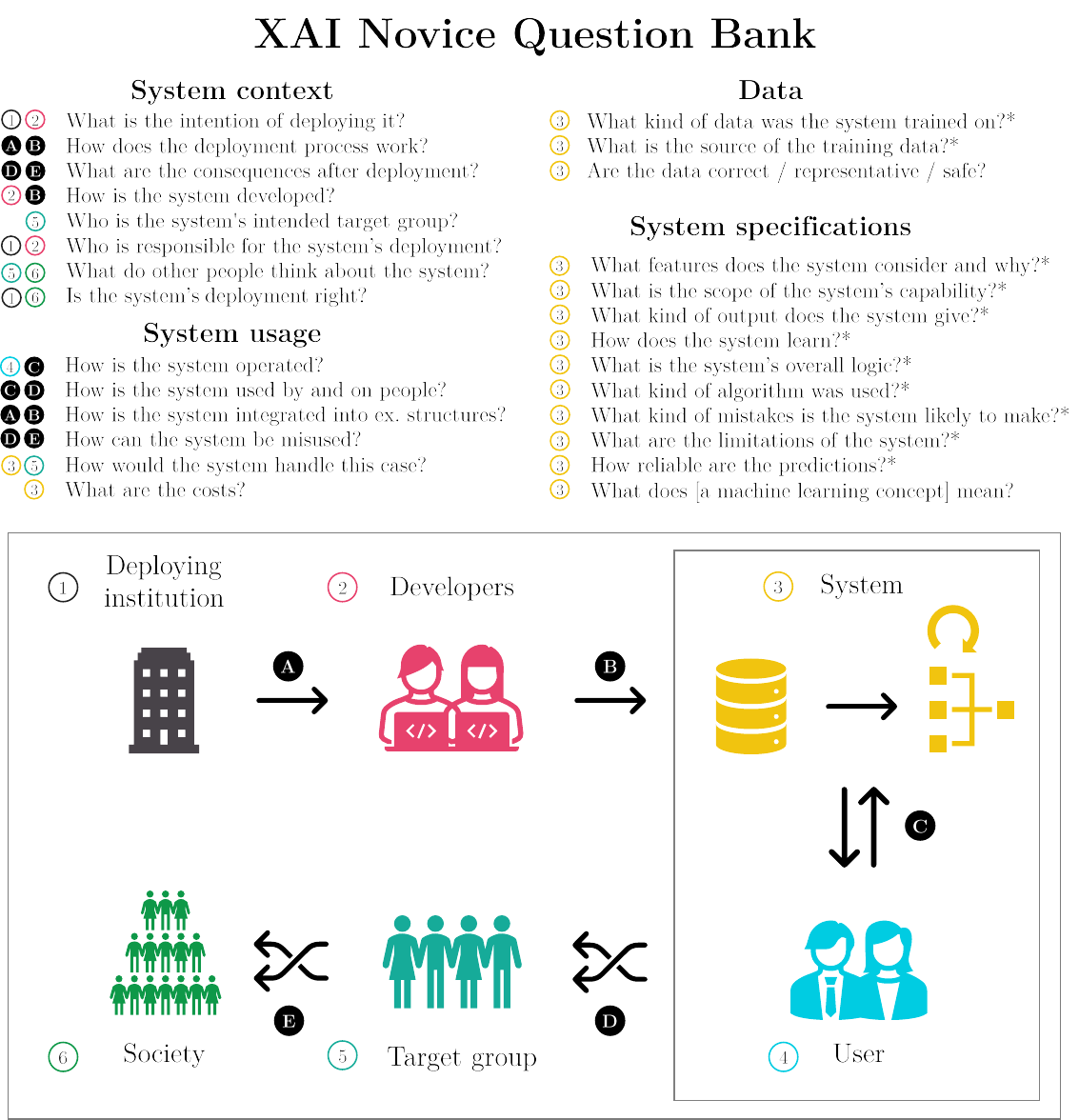}
    \caption[XAI Novice Question Bank]{\textbf{The XAI Novice Question Bank and system-inquiry diagram.} Depicted above are four categories of questions that subsume participants' inquiries about the two ADM use cases (Section~\ref{sec:use_cases}). An asterisk (*) indicates that the question is already present in the XAI Question Bank~\cite{liao2020}. Numbers and letters are added to the side of each question to refer to stakeholders (pictograms) and procedures (arrows) in the system deployment process shown below. System and user are framed together to indicate this part of deployment as the core interaction with the system. In the system section, the arrow indicates the separation of data and model. The two-way arrows in procedure C depict interactions between the user and the system, and the twisted arrows in procedures D and E indicate the complex effects that system deployment and usage have on both the target group and society.}
    \label{fig:xainqb}
\end{figure}

In this section, we present our results, structured according to our research questions: information needs of AI novices (RQ1, Section~\ref{sec:rq1}), participant understanding and decision confidence (RQ2, Section~\ref{sec:rq2}), and perceived risks and benefits (RQ3, Section~\ref{sec:rq3}).

\begin{table}[]
\small
\caption{\textbf{XAI Novice Question Bank: Tabular view.} Collection of participants' questions as depicted in~\autoref{fig:xainqb} and the type of information provided in response. An asterisk (*) indicates that the question is already present in the XAI Question Bank~\cite{liao2020}.}
\resizebox{\textwidth}{!}{%
\begin{tabular}{p{1.7cm}|p{2.2cm}p{6.2cm}p{6.2cm}}
\textbf{Category} & \textbf{Subcategory} & \textbf{Question} & \textbf{Information provided} \\ \hline
 
\textbf{System context} & Intention & What is the intention of deploying it? & Design idea and underlying   motivation \\
 & \begin{tabular}[t]{@{}l@{}}Deployment \\      process\end{tabular} & How does the deployment process work? & Process of preparing the system   for operation \\
 & Consequences & What are the consequences after deployment? & 
 Possible effects of deployment \\
 & Development & How is the system developed? & Development process and   institution \\
 & \begin{tabular}[t]{@{}l@{}}Target \\      group\end{tabular} & Who is the system's intended target group? & Scope and features of the target   group \\
 & Responsibility & Who is responsible for the system’s deployment? & Responsible actors \\
 & Ethical considerations & Is the system's deployment right? & None, answer relies on personal assessment \\ \hline
 
\textbf{System usage} & Tool & How is the system operated? & Usage in practice, everyday application \\
 & \begin{tabular}[t]{@{}l@{}}Sociotechnical \\      component\end{tabular} & How will the system impact interpersonal relations? & Interpersonal work, substitution of people \\
 & \begin{tabular}[t]{@{}l@{}}Part of an \\      organization\end{tabular} & How is the system integrated into existing structures? & Interfaces, processes, organization's   influence \\
 & Misuse & How can the system be misused? & Scenarios of misuse \\
 & Example & How would the system handle [this case]? & Specific example walkthrough \\
 & Costs & What are the costs? & Financing, personal costs \\ \hline

\textbf{Data} & Type & What kind of data was the system trained on?* & Content and structure of   training data \\
& Source & What is the source of the training data?* & Collection and  storage of training data \\
& Attributes & Are the data correct / representative / safe? & Quality, biases, safety of   training data \\ \hline
 
\textbf{System specs.} & Features & What features does the system consider and why?* & Information fed into system \\
 & Functionality & What is the scope of the system’s capability?* & Supported range of functions \\
 & Output & What kind of output does the system give?* & Information received from system \\
 & Learning & How does the system learn?* & Automated improvement of system \\
 & Logic & What is the system’s overall logic?* & General process overview \\
 & Algorithm & What kind of algorithm was used?* & Specific model used \\
 & Errors & What kind of mistakes is the system likely to make?* & Expected errors \\
 & Limitations & What are the limitations of the system?* & Known boundaries \\
 & Reliability & How reliable are the predictions?* & Dependability of output \\
 & Reflexive & What does {[}a machine learning concept{]} mean? & ML concept information
\end{tabular}}
\label{fig:xainqb_tabular}
\end{table}

\subsection{RQ1-Information Needs: What information do AI novices who could
be affected by algorithmic decisions need to decide about
adopting an ADM system?}
\label{sec:rq1}

To address RQ1, we coded participants' questions raised in the inquiry phases to identify reoccurring question categories. The found categories are \textit{system context}, \textit{system usage}, \textit{data}, and \textit{system specifications}, and are each composed of several prototypical questions. An overview is provided in~\autoref{fig:xainqb}, a tabular representation in~\autoref{fig:xainqb_tabular}. In the following, each of the four question categories is described and split into subcategories. We denote the total number of questions asked in each subcategory in brackets.

\subsubsection{\textbf{System context}: Intention, deployment process, consequences, development, target group, responsibility, and ethical considerations.} This category captures questions that elicit background information on the system's development and deployment (69 questions total).  

\textbf{Intention: What is the intention of deploying it?} Questions about the intention of the system's deployment were asked for both use cases (18 total), with a majority in employment prediction. These questions aimed at the underlying reasoning for introducing the system. Here we differentiate i) practical intention, such as data collection, "\textit{How did they come up with the idea of GPS? What's the intention behind it?}" (B4); and ii) structural intention, such as societal benefit, "\textit{What is the benefit for society? What did they expect from this algorithm?}" (A1).

\textbf{Deployment process: How does the deployment process work?} Questions about the general deployment process were primarily asked in the employment prediction use case (13 of 19 total). This includes inquiries about how the system would be tested and how comparable systems were deployed in other countries.

\textbf{Consequences: What are the consequences after deployment?} Questions on what would follow from the system's deployment were asked in low counts in both use cases (9 total). Topics of inquiry included the effect on the available workforce and care ratio and tended to focus on risks and challenges:

\begin{quote}
    \textit{Won't I lose labor potential if I only decide based on educational background and so on? If this algorithm is introduced, will it not lead to the loss of workforce?} (A1)
\end{quote}

\textbf{Development: How is the system developed?} Questions on the system's technical development emerged mainly in the employment prediction use case (8 of 9 total). They covered, e.g., the contracted company, duration of development, and the collaboration between developers and domain experts. Personnel counselors primarily asked these questions, who related the information to their experience with contracted software developers and compared it to known procedures.  

\textbf{Target group: Who is the system's intended target group?} Questions on the system's target groups emerged mainly in the wristband use case (8 of 10 total) and covered, e.g., the target group's physical mobility and financial situation. In contrast, the two inquiries in the employment prediction use case considered target groups more on a societal level: "\textit{Would the system be applied for every population group? Can we say that every citizen would then be covered?}" (A11).

\textbf{Responsibility: Who is responsible for the system's deployment?} These questions (4 total) addressed responsibility for the system's deployment, including the political decision, financing, project management, and technical implementation. 

\textbf{Ethical considerations: Is the system's deployment right?} Questions in this subcategory addressed perceived social wrongs or difficult trade-offs on personal and societal levels, which might be aggravated by the ADM system's introduction (9 total). Personal topics included perceptions of unfairly distributed resources, perceived inaccessibility to own data, and over-reliance on care personnel for patients' safety ("\textit{If I know that the person falls all the time, what do I do? If I let them walk around anyway, am I then responsible?}" (B1)). Societal topics included the (in)sensibleness of sanctioning job-seekers, reflections on whether an ADM system could be more profitable if it did not aim for people's well-being and the (in)validity of quantifying people's attributes ("\textit{Do you believe that you can categorize every human being in such a scientific way? By some gradations and variables?}" (A4)).        

\subsubsection{\textbf{System usage}: As a tool, as a sociotechnical component, as part of an organization, misuse, examples, costs.} This category captures participants' information needs about how the system would be used in practice and as a concrete application. Questions on usage were asked frequently in both use cases (89 questions in total by over 20 participants). 

\textbf{Usage as a tool: How is the system operated?} Questions on practical usage details were almost exclusively gathered in the wristband use case (30 of 31 total). These typically concern how the wristband would be used, including waterproofing, charging, operation, etc. Questions like these would likely be covered by a manual accompanying the wristband. 

\textbf{Usage as a sociotechnical component: How will the system impact interpersonal relations?} Questions in this category asked who would use the system and how the system would influence relations between people. These were gathered mostly in the employment prediction use case (13 of 14 total). Participants' inquiries focused on how the system would affect interaction between counselor and job-seeker and covered both the counselor's and job-seeker's personal experience as well as their relationship:

\begin{quote}
    \textit{Some people open up only after the second, third, or fourth conversation. [...] Here, it's really about subjective perception. What do I learn from the interview? What do I learn from the person? And what do I make of it? With the algorithm, I would need to know that this dialogue is still possible.} (A2)
\end{quote}

\textbf{Usage as part of an organization: How is the system integrated into existing structures?} Questions that focused on the system's interaction with existing organizational structures were present in both use cases (28 total) and covered a wide range of topics, including data processing, data access, information flow, and contestability: "\textit{What are the options for correction? Is there a system loop that handles this?}" (A7).
    
\textbf{Misuse: How can the system be misused?} A noticeably low number of questions (2 total) addressed the potential of misusing the system; both emerged in the employment prediction use case. While one participant asked about using the system to commit welfare fraud, the other was concerned about the system being used by the police.

\textbf{Examples: How would the system handle [this case]?} Questions in this subcategory aimed to examine how the system would behave given a specific example case and emerged in low count in both use cases (6 total). Specifically, several participants were interested in how the algorithm would handle their case.

\textbf{Costs: What are the costs?} Questions about costs pertained to both personal costs ("\textit{What would it cost me?}" (B5)) as well as organizational expenses ("\textit{If the state pays, what is the cost? Is the money well invested?}" (A4)). This subcategory appeared more often in the wristband use case (6 of 8 total).

\subsubsection{\textbf{Data}: Type, source, attributes.} This category captures questions about the data used by the system. These questions emerged in both use cases (43 questions total) and mostly occurred only after participants received the XAI Question Bank~\cite{liao2020} as reference. The question bank already contains questions about \textbf{data type} (\textit{What kind of data was the system trained on?}) and \textbf{data source} (\textit{What is the source of the training data?}). In our study, multiple participants also asked about data privacy and misuse (11 questions by 8 participants). Therefore, we extend the data category with a subcategory on \textbf{data safety} (\textit{Are the data correct / representative / safe?}).  

\subsubsection{\textbf{System specifications}: Features, functionality, performance, output, predictions, logic, reflexive inquiry.} This category covers all questions focusing on specific aspects of the ADM system and has the highest question count of all categories (187 total). In contrast to the previous categories, many questions on system information emerged only after participants were provided with the XAI Question Bank~\cite{liao2020} in the second inquiry phase. We focus on describing questions about the system's features, functionality, performance, and reflexive inquiry, as the remaining subcategories mostly contain questions adopted verbatim from the XAI Question Bank.

\textbf{Features: What features does the system consider and why?} Questions on the system's input features emerged almost exclusively in the employment prediction use case (28 of 31 total). Most questions on features emerged before participants were given the XAI Question Bank~\cite{liao2020} but mirrored questions covered in the bank, e.g., \textit{Which features are considered?}, \textit{How is a certain feature weighed?}, and \textit{Why is this feature included?} This points to shared information needs between AI novices and experts, as questions on features in the question bank were sourced from stakeholders \textit{with} technical knowledge. However, the use case appears to have a strong impact on whether features are inquired at all.    

\textbf{Functionality: What is the scope of the system? Can it do...?} Questions on the system's functionality emerged almost exclusively in the wristband use case (30 of 34 total). The 'functionality' subcategory is close to the 'usage as a tool' subcategory but differs in its focus: Questions in this subcategory asked about functional scope in general ("\textit{How does the wristband work?}" (B5)), whereas questions in the "usage as a tool" subcategory" instead asked how the system would be used specifically and practically ("\textit{Does that mean you can also shower with it?}" (B4)). 

\textbf{Errors, limitations, reliability} Questions about errors, limitations, and reliability (21, 23, and 7; 51 total) emerged in both use cases equally and covered topics such as error frequency and susceptibility, handling incomplete data, constraints of usage scenarios, and precision. While questions on reliability only emerged after reference to the bank, questions on limitations and errors emerged organically even in the first inquiry phase: "\textit{I mean, the question is, if the system really does make mistakes, how quickly can you give the feedback: 'Oops, that wasn't a fall, everything's okay.' Is that possible?}" (B4).

\textbf{Reflexive inquiry: What does [a machine learning concept] mean?} This subcategory captures questions about the meaning of certain notions in machine learning, which emerged in both use cases (14 total) after participants read through the XAI Question Bank~\cite{liao2020}. Questions were often prompted by terms such as \textit{algorithm}, \textit{biases}, and \textit{predictions}. However, a lack of understanding in vocabulary often correlated with a lack of understanding of the corresponding concept: "\textit{'What kind of data was the system trained on?' I don't understand the 'trained' part. That's just an input, isn't it?}" (A3). We thus adapt the original bank's question slightly such that it captures questions about machine learning \textit{concepts} instead of \textit{terminologies}.  

\subsection{RQ2-Understanding: How do question-driven explanations interact with participants' understanding?}
\label{sec:rq2}

This section describes how the explanations given in response to participants' questions interacted with their understanding and decision confidence. First, we focus on the effects of question-driven explanations on self-reported understanding and decision confidence (Section~\ref{sec:sru}). Second, we describe how participants' questions can be analyzed using the framework of "explanatory stances" and which challenges in understanding arose (Section~\ref{sec:und->inq}). 

\subsubsection{\textbf{How question-driven explanations impact understanding and confidence.}}

Participants reported their perceived understanding and decision confidence on 5-point Likert scales before and after the inquiry phases, i.e., before and after receiving explanations (overview depicted in~\autoref{fig:self_reported_und_table}).\footnote{One participant decided not to vote on the use case and did not answer the second query out of concern about how the study results would be used.} We describe three main aspects of these self-reports: participants' overall perceived understanding, how their understanding changed throughout the study, and how their understanding interacted with their decision confidence. 
\label{sec:sru}

\begin{table}[h]
\caption{\textbf{Change in self-reported understanding and decision confidence.} Participants indicated their perceived understanding of the system and confidence in their decision on a 5-point Likert scale before and after receiving question-driven explanations in the inquiry phases (Section~\ref{sec:Study_setup}). This table summarizes the change between their two self-reports. Notably, changes in perceived understanding did not necessarily lead to increases and decreases in confidence. Changes between reports are encoded with symbols and colors, + ({\color[HTML]{548235}{increase}}), - ({\color[HTML]{FF0000} {decrease}}), and = ({\color[HTML]{DEA900} {no change}}). Columns indicate participant ID and use case.}
\resizebox{\textwidth}{!}{
\begin{tabular}{lcccccccccccc}

\textbf{Employment prediction use case} & \textbf{A6} & \textbf{A11} & \textbf{A4} & \textbf{A1} & \textbf{A8} & \textbf{A3} & \textbf{A10} & \textbf{A5} & \textbf{A2} & \textbf{A12} & \textbf{A9} & \textbf{A7} \\

$\Delta$ Self-reported understanding & {\color[HTML]{548235} \textbf{++}} & {\color[HTML]{548235} \textbf{+}} & {\color[HTML]{DEA900} \textbf{=}} & {\color[HTML]{DEA900} \textbf{=}} & {\color[HTML]{DEA900} \textbf{=}} & {\color[HTML]{DEA900} \textbf{=}} & {\color[HTML]{DEA900} \textbf{=}} & {\color[HTML]{DEA900} \textbf{=}} & {\color[HTML]{FF0000} \textbf{-}} & {\color[HTML]{FF0000} \textbf{-}} & {\color[HTML]{C00000} \textbf{- - -}} & \textbf{n/a} \\ \hline

$\Delta$ Confidence & {\color[HTML]{548235} \textbf{+}} & {\color[HTML]{DEA900} \textbf{=}} & {\color[HTML]{548235} \textbf{+}} & {\color[HTML]{548235} \textbf{+}} & {\color[HTML]{DEA900} \textbf{=}} & {\color[HTML]{DEA900} \textbf{=}} & {\color[HTML]{DEA900} \textbf{=}} & {\color[HTML]{FF0000} \textbf{- -}} & {\color[HTML]{DEA900} \textbf{=}} & {\color[HTML]{DEA900} \textbf{=}} & {\color[HTML]{DEA900} \textbf{=}} & \textbf{n/a} \\

 &  &  &  &  &  &  &  &  &  &  &  &  \\
\textbf{Health wristband use case} & \textbf{B10} & \textbf{B8} & \textbf{B7} & \textbf{B2} & \textbf{B4} & \textbf{B5} & \textbf{B6} & \textbf{B9} & \textbf{B12} & \textbf{B1} & \textbf{B3} & \textbf{B11} \\

$\Delta$ Self-reported understanding & {\color[HTML]{548235} \textbf{+}} & {\color[HTML]{548235} \textbf{+}} & {\color[HTML]{DEA900} \textbf{=}} & {\color[HTML]{DEA900} \textbf{=}} & {\color[HTML]{DEA900} \textbf{=}} & {\color[HTML]{DEA900} \textbf{=}} & {\color[HTML]{DEA900} \textbf{=}} & {\color[HTML]{DEA900} \textbf{=}} & {\color[HTML]{DEA900} \textbf{=}} & {\color[HTML]{FF0000} \textbf{-}} & {\color[HTML]{FF0000} \textbf{-}} & {\color[HTML]{FF0000} \textbf{-}} \\ \hline

$\Delta$ Confidence & {\color[HTML]{548235} \textbf{+}} & {\color[HTML]{FF0000} \textbf{-}} & {\color[HTML]{548235} \textbf{++}} & {\color[HTML]{548235} \textbf{+}} & {\color[HTML]{DEA900} \textbf{=}} & {\color[HTML]{DEA900} \textbf{=}} & {\color[HTML]{DEA900} \textbf{=}} & {\color[HTML]{DEA900} \textbf{=}} & {\color[HTML]{FF0000} \textbf{-}} & {\color[HTML]{548235} \textbf{++}} & {\color[HTML]{548235} \textbf{+}} & {\color[HTML]{548235} \textbf{+}}
\end{tabular}}
\label{fig:self_reported_und_table}
\end{table}

\textbf{Self-reported understanding was high overall, but participants define 'high understanding' differently.} In the first self-reported understanding, before receiving explanations, only 4 of 24 participants indicated that they did not understand the ADM system well. In the second self-report, after receiving explanations, there even was only 1 participant who indicated that they did not understand the system and who perceived a substantial decrease in their understanding ("\textit{Because we talked, I realized that I don't understand it.}" (A9)). Multiple participants experienced this decrease (A2, A12, B1, B3, B11).

While the overall self-reported understanding was, therefore, high, the level of detail in their explanations for these reports differed noticeably. For example, domain expert A2 explained: "\textit{The system should aid in deciding which measure to offer to a job-seeker. [...] The question is how the algorithm is created, but the basic classification, I think, makes sense.} (A2). In contrast, some participants explained their perceived high understanding with more general observations, such as their reception of AI in the media: "\textit{There is a lot of talk about artificial intelligence in the media. Mostly at night. I watch it when I can’t sleep.}" (B5). Other participants spoke about the general employment market (A5), the consequences of digitization (A9), and comparable systems (e.g., B8, A12). This illustrates a discrepancy in what people define as "high" understanding, possibly resulting from the difficulty of assessing one’s understanding~\cite{keil2006} (Section~\ref{sec:discussion}). 

\textbf{Question-driven explanations mainly did not influence self-\\reported understanding.} After proceeding through the inquiry phases, most participants reported their understanding to have remained the same; 4 participants reported increased understanding, and 6 reported declines (as seen in~\autoref{fig:self_reported_und_table}). As expected, increased self-reported understanding was explained through information gained in the inquiry phases. In contrast, decreases in understanding seemed to have various reasons, e.g., for A2, the decrease in understanding was accompanied by heightened scrutiny:

\begin{quote}
    \textit{Interesting, I was so convinced about it at first, but now there's something missing from my point of view. Basically, I'm still convinced that it's a good thing, but it's not ready yet.} (A2)
\end{quote}

Previous studies have observed that information gain can paradoxically result in decreased perceived understanding, even though objective understanding increased, likely resulting from a new awareness for unknown information~\cite{cheng_explaining_2019}. 

\textbf{Confidence increased for 9 participants but seems not to depend on understanding.} Compared to understanding, question-driven explanations increased participants' decision confidence more often (as seen in~\autoref{fig:self_reported_und_table}). Several participants noted that it would increase their confidence further if they saw the system in practice: \textit{"By and large, I know how the system works. Otherwise, to be sure, I would have to see how it actually works in practice."} (A4). Participants, however, offered little explanation as to why understanding did not increase together with their confidence.   

\subsubsection{\textbf{How question-driven explanations interacted with explanatory stances and which challenges arose.}} 

The cognitive sciences outline several processes governing how people seek and process information (described in Section~\ref{sec:related_work}). In the following, we analyze two main aspects of participant understanding: the role of explanatory stances and the challenges that prevented participants from acquiring a "working understanding"~\cite{keil2006}.
\label{sec:und->inq}

\textbf{Explanatory stances offer a way of matching explanations to participants' information needs.} Explanatory stances are a way of describing how people acquire information to develop a predictive strategy of a system's (or person's) behavior~\cite{dennett_intentional_1998}. To this end, people can take different strategies, which might be focused on a system's physical attributes, on its design and functions, or, if those are inaccessible, on the "beliefs [an] agent ought to have given its place in the world and its context"~\cite{dennett_intentional_1998}, which define its intentions and actions.    

To analyze if participants preferred a specific strategy of seeking information, we apply the concept of the three explanatory stances (mechanical, design, and intentional) to their questions during the inquiry phases.\footnote{An introduction is provided in~\autoref{sec:related_work}.} 
We conduct this analysis by assigning participants' questions the likeliest of the three stances. This assignment is an interpretation of their chosen stance derived from the question's phrasing and the conversation's context. Consequently, several questions might be assigned to more than one stance, depending on the interpretation of the information needs (discussed in Section~\ref{sec:discussion_stances}). In the following, we describe our analysis of which stances participants applied for their questions about the ADM systems:

\begin{itemize}
    \item \textit{Questions about \textbf{data} and \textbf{system specifications} often applied the mechanical stance.} The mechanical stance focuses on parts, processes, and mechanisms and does not consider design or intentions~\cite{lombrozo_mechanistic_2019}. Participants tended to apply this approach to learn about the technical details of the system, such as the data basis, features, predictions, and output. This direction of inquiry often resulted from reference to the XAI Question Bank~\cite{liao2020}, which contains many questions about technical details, including "\textit{What is the sample size of the training data?}" (A4), "\textit{What are the features with the evaluation criteria?}" (A7), and "\textit{How sensitive is it?}" (B8). However, small differences in question phrasing can lead to different explanations. Compare: \textit{"What is the sample size of the training data?"} (A4) and \textit{"The sample size of the data, I assume, is large enough?"} (A5). While the first one is taken verbatim from the XAI Question Bank~\cite{liao2020} and can be said to focus on a "part" of the system (the training data), the second raises the question of what the sample size should be \textit{large enough for}, which rather applies the design stance by asking about functional information.  
    
    \item \textit{Questions about the system's \textbf{usage} and \textbf{context} applied the design stance.} The design stance focuses on how functions and goals are realized in terms of design~\cite{keil2006}. Therefore, we assign participants' questions about the system's design's practical and conceptual aspects to this stance. Examples of practical design aspects include questions about usage details and operation instructions ("\textit{Do you wear it day and night?}" (B1)). In contrast, examples of conceptual design aspects include questions on the motivation behind certain functionalities (\textit{"What's the benefit of knowing whether the patient was at the refrigerator at 3:00?"} (B4)). Further, we assign questions about context information to this stance when they acquire information about, e.g., the characteristics of the target group ("\textit{This thing is only supposed to be distributed to care home residents, right?}" (B12)) or control mechanisms ("\textit{What options for correction are there?}" (A7)). 
    
    \item \textit{Questions about \textbf{context} and \textbf{values} applied the intentional stance.} The intentional stance describes a system's behavior based on beliefs and motives embedded in it, which are assumed (in the absence of mechanical and design information) to drive its actions~\cite{ dennett_intentional_1998}. We assigned participants' questions to this stance when they addressed topics such as the intentions of the deploying institutions ("\textit{Would that mean that the job counselors would need to perform better?}" (A4)) and the rights of affected persons ("\textit{If I'm analyzed, I would like to have the results. Is there such a thing as a right of appeal?}" (A8)). Further, the intentional stance also tended to be used when participants suspected that their values would not be considered in the system (\textit{"When this comes, who guarantees that it is not forced on you?" (B6))}. Notably, many questions in the intentional stance focused on the beliefs and values of the deploying institution,  which connects to literature emphasizing the importance of people's perceptions toward the institution~\cite{cavaliere_poisons_2022}. This showed that participants regarded the job agency critically and is reflected in participants' perceptions of the systems' risks and benefits (Section~\ref{sec:rq3}).  
\end{itemize}

\textbf{Challenges in acquiring information to develop understanding.} Explanations bring the incompleteness of one's understanding into focus and prompt specific reactions: filling the gap with further details, "outsourcing" it to another mind, or mentally rendering it irrelevant~\cite{keil2006}. The explanation setting used in this study explicitly supported filling gaps with information by letting participants flexibly inquire about any topic. However, participants also met challenges in acquiring information and developing understanding. In the following, we describe three of these challenges:   
\label{sec:understanding_process} 

\begin{itemize}
    \item \textit{\textbf{Blind spots inhibit further inquiry}.} In the closing interviews, multiple participants expressed that they did not understand all relevant aspects of the system but could not articulate which information was missing. Reasons given for this effect included a lack of mathematical knowledge (A6), too little preparation (A8), pressure of the study situation (B5, A12), and uncertainty about the system's practical behavior (A1, A4, B1, B9): "\textit{Experience has taught me that it's always the practical application that raises issues.}" (B1). We call understanding gaps that could not be closed "blind spots".
    \item \textit{\textbf{Outsourcing understanding divides cognitive labor} but risks over-dependence.} Multiple participants asked about other people's judgment of the system, e.g., the job center's employees (A8), digital natives (A9), their families (B11), and the study examiner (A11). This "outsourcing"~\cite{keil_folkscience_2003} is a common process to divide the cognitive labor of understanding between trusted people. Critically, some participants stated intention to outsource their full understanding: \textit{"The first question is: Would you have said yes to this system?"} (A9). Trust is thus a crucial part of stakeholders' understanding when outsourcing is used, especially as it risks over-dependence on the entrusted person.
    \item \textit{\textbf{Abandoning of understanding} undermines the explanation effort.} Some participants expressed that they would likely not care about the system even if it was used on them, resulting from a lack of interest or the belief that their voice would be irrelevant: \textit{"I still think that it's not my interest that is being represented here, but actually the interest of the job agency."} (A8). A pronounced disinterest in topics despite their relevance for oneself is known in political science as "civic ignorance"~\cite{lupia_uninformed_2016}\footnote{The term describes that ignorance on public matters applies to almost everyone~\cite{lupia_uninformed_2016}.}. Research has been conducted on ways to remedy this disinterest and will be briefly discussed in Section~\ref{sec:discussion}.
\end{itemize}

\subsection{RQ3-Perception: How do question-driven explanations impact perceptions of the systems' risks and benefits?}
\label{sec:rq3}

\begin{figure*}[]
    \includegraphics[width=\textwidth]{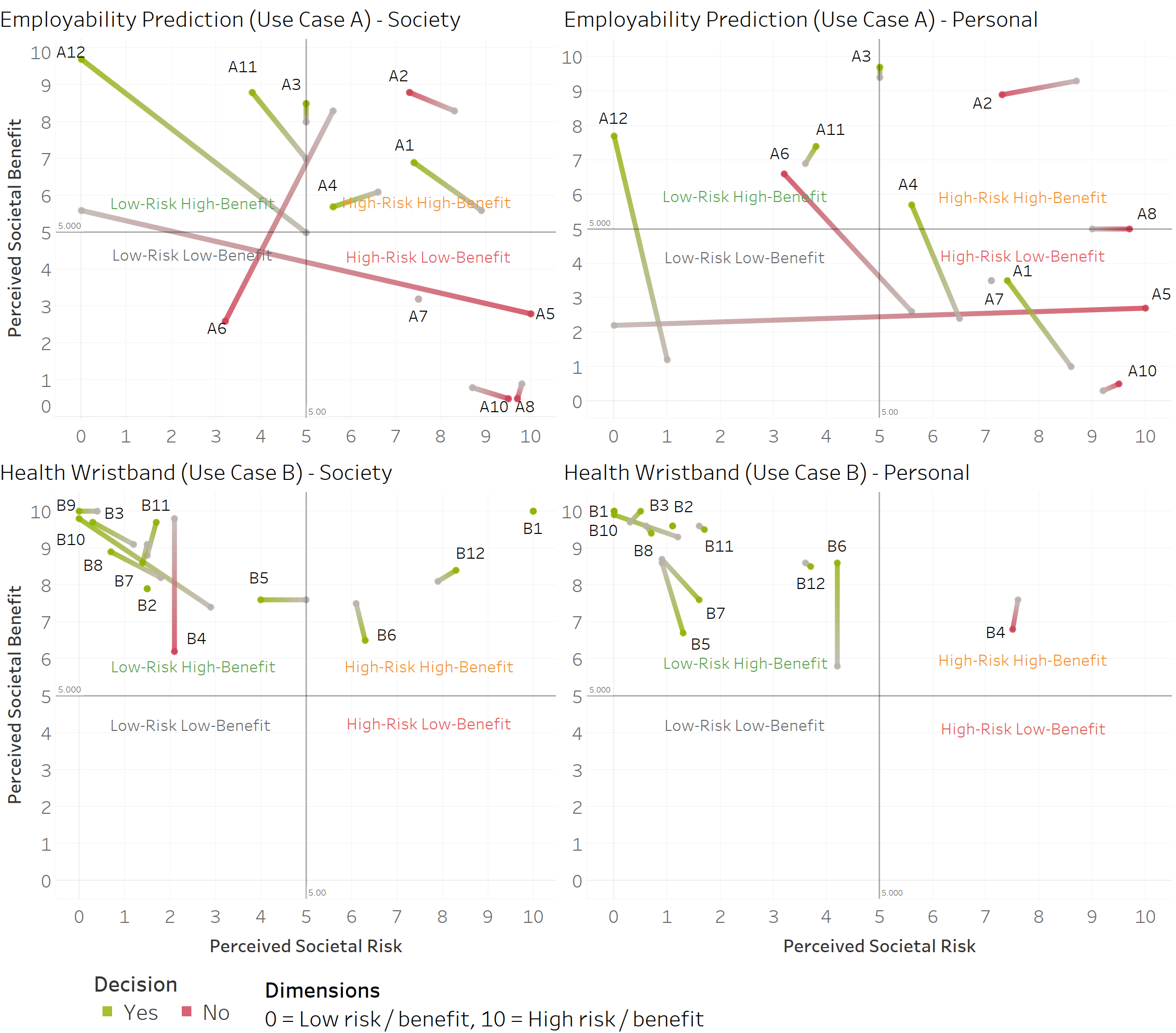}
    \caption[Risk and benefits]{\textbf{Changes in perceived risks and benefits.} Participants were asked two times for their perceptions of the given ADM system's risks and benefits for both society and them personally (as described in Section~\ref{sec:method}) on scales from 0 (low) to 10 (high). Lines depict perception changes from before (gray) to after (colored) the inquiry phases, split between the employment prediction use case (A) above and the health wristband use case (B) below. While perceptions changed drastically in A and are distributed throughout all quadrants, perceptions in B changed comparatively little and remained mostly in the upper left quadrant.}
    \label{fig:r&b_society}
\end{figure*}

Participants were asked how they perceived the risks and benefits of the ADM system before receiving explanations and after making the deployment decision. The two-dimensional grids used to elicit these perceptions divide the space into four \textit{quadrants}, corresponding to four combinations of risk and benefits (low-risk low-benefits, low-risk high-benefits, high-risk high-benefits, and high-risk low-benefits; Section~\ref{sec:method_self_reports}). Changes in their perceptions between the first and second report are visualized in~\autoref{fig:r&b_society}, showing solid changes in the employment prediction use case and lesser changes in the health wristband use case. In the following, we describe participants' perceptions in both use cases and outline potential reasons for the changes. 

\newpage

\textbf{Perceptions changed enormously in employment prediction use case}. Participants in the employment prediction use case showed very different initial perceptions of the system, ranging over all combinations of risks and benefits and showing little focus on specific attitudes (as seen in~\autoref{fig:r&b_society}). Further, participants enormously changed their perceptions of the ADM systems after receiving explanations, often transitioning into other perception quadrants due to newfound information. 

\begin{quote}
    A6: \textit{What does "regional labor market" mean? That a large number of jobs are registered as vacant?} \\
    Examiner: \textit{Yes, the city was divided into regional labor market types. It was intended to show how many jobs are available, but it is also a placeholder variable for the place of residence.} \\
    A6: \textit{I find that quite discriminatory, I have to say.}
\end{quote}

Frequently mentioned benefits included a potentially reduced workload, reduced administrative overhead, and guidance for personnel counselors, while frequently mentioned risks included a lack of care for the individual, fear of discrimination, and overreliance on algorithmic decisions. A11, for example, reported an increase in perceived societal benefit, explaining:

\begin{quote}
    \textit{I believe that the state could save money in the broadest sense. And this money could be used to plug other holes. [...] It would only be a risk because soft skills and the emotional part are not included. That's the only risk I can see.} (A11)
\end{quote}

In contrast, A2 reported a slight decrease in perceived risk and varying changes in benefit while also voting "no", stating:

\begin{quote}
    \textit{What I'm missing here is that no consideration is given to whether the person looking for work can continue in their core business or needs a new career path. [...] I miss the jobseeker's personal perspective. The hard facts are the smaller issue for me.} (A2)
\end{quote}

We observed that question-driven explanations had mixed effects on participants' perceptions in the employment prediction use case. While the provided information indeed enabled a shift in perspective, the trajectory of that shift seems to differ heavily, even though the information provided did not change between participants. 

\textbf{Perceptions changed less in the health wristband use case}. Participants in the health wristband use case saw the system clearly as having low risk and high benefit. Changes in perceptions were minor and never resulted in transitions between perception quadrants, showing that participants perceived the system positively before and after receiving explanations. 

\begin{quote}
      \textit{Through the question and answer sessions, my idea of the wristband has become more concrete. I realized that the thing can do other things than I had imagined. But with risks and benefits, I'm sticking with what I had before, that hasn't changed.} (B7)
\end{quote}

For some participants, the wristband presented a trade-off between the values of safety and privacy, which they weighed against each other. Despite little overall change in perceptions, this highlights that participants considered the risk of surveillance and the benefit of the wearer's safety and came to different conclusions. B11, for example, personally did not see any issues, while B4 specifically mentioned the feeling of being surveyed:
\label{sec:r&b_B}

\begin{quote}
    \textit{For me, I don't see any disadvantages. But I could imagine that there are people who say: "I don't want to be monitored. Even if I fall down and nobody comes and I bleed to death, then I'll just bleed to death."} (B11) \\
    \textit{I would rate the risk as high because I believe it's going in the direction of surveillance.} (B4)
\end{quote}

In summary, participants thus weighed the given information and decided based on it, which suggests that the explanations fulfilled their purpose in revealing potential value conflicts in the system's deployment. Making this conflict explicit, B12 stated:

\begin{quote}
    \textit{The risk of data collection is a risk inherent to the system. Either you accept this risk and have the benefit of these systems, or you do without it. You probably can't have one without the other.} (B12)
\end{quote}

\newpage
\section{Discussion}
\label{sec:discussion}

This section outlines the theoretical and practical implications of our study. It addresses AI novices' information needs, the interaction between question-driven explanations and understanding, and the inclusion of domain experts and decision subjects in explanation design. These implications are summarized at the end (Section~\ref{sec:discussion_implications}).

\subsection{Theoretical implications}

\subsubsection{Designing explanations for AI novices}
\label{sec:discussion_information}

One of XAI's enduring key questions is: What makes a "good" explanation? Recent work outlines that, to produce understanding, explanations should be contrastive, selected, and social while being first and foremost \textit{contextual}~\cite{miller_explanation_2019}, that they should interface directly with people's causal reasoning~\cite{byrne_good_2023} and consider the learner's aim and prior knowledge~\cite{mueller2019explanation}. Fulfilling these requirements puts a high demand on explanation design. Further, existing explanation approaches do not typically address the needs of AI novices~\cite{shin_algorithms_2023}, even though they compose multiple stakeholder groups, including people affected by algorithmic decisions and end-users. We thus argue that one step in designing explanations for AI novices is to identify their information needs and examine how they depend on participants' perceptions, stakeholder roles, and domain expertise. In the following, we present how the findings of our study support this design approach. 

Our findings show that \textbf{AI novices' information needs cover all parts of an ADM system's "lifecycle"}~\cite{dhanorkar_who_2021}, from technical details over deployment consequences to different dimensions of usage. Our proposed "XAI Novice Question Bank" (\autoref{fig:xainqb}) provides a summary of these topics, which can inform future explanation design. However, we also observe differences in information needs between use cases and differences to~\citet{liao2020}'s study with AI practitioners. This is in line with previous work, which shows that various contextual factors mediate which information needs participants articulate, including familiarity with the presented system~\cite{kramer2018}, tangibility of the system~\cite{long_role_2021}, perceived risks and benefits~\cite{Araujo2020} (e.g., trade-off between safety and privacy in the wristband use case), domain expertise~\cite{wang21}, and technical expertise~\cite{cheng_explaining_2019}. 

\subsubsection{Anticipating and handling explanations of intention.}
\label{sec:discussion_understanding}

Explanations have the purpose of enabling stakeholders to attain their aims~\cite{langer_what_2021}, such as facilitating the evaluation of an ADM system's values~\cite{SHIN2021, shin_algorithms_2023}, or performing an action, like contesting a decision or requesting human intervention~\cite{Alfrink2022}. In our study, participants were asked to vote over adopting the ADM systems to incentivize information acquisition through questions. I.e., the "purpose"~\cite{freiesleben2023} of the explanations was to enable a well-informed decision. For several participants, this succeeded, as they used the explanations to underpin their decision and reported a slight increase in decision confidence (\autoref{fig:self_reported_und_table}). 

At the same time, as participants' information needs differed between use cases, their perception of what a well-informed decision would need to consider differed. Participants tended to inquire more about societal and structural factors in the employment prediction use case and more about practical and operational aspects in the wristband use case. Framed in the context of the explanatory stances, many participants used the "intentional stance"~\cite{dennett_intentional_1998} to ask questions about the \amsalgorithm and the beliefs and values it represents (Section~\ref{sec:rq2}). This demand for explanations of intention was likely due to the system's close association with the job agency, as previous work has shown that the reputation of the deploying institution can affect perceptions of ADM systems~\cite{brown_toward_2019, woodruff_qualitative_2018, lee_webuildai_2019}. 

However, delivering explanations of the intention behind a system's deployment is difficult, as it means interpreting the values involved and choosing how to present them -- a process that rather should be performed by each participant individually due to its inherent subjectivity. Further, information about the true purpose of an ADM system's deployment can be hard to attain, as it is often part of a deploying institution's non-public decision-making process and can be kept intentionally opaque~\cite{Rudin2019}. 

Consequently, \textbf{whether explanations of intention should be included needs to be matched to the explanation's purpose}. For example, if the explanation's purpose is solely a mechanical or functional description of the system to enable operation, intentional information is likely negligible. In contrast, when stakeholders should form value assessments or investigate possibilities for contestation~\cite{Alfrink2022}, intentional explanations can contribute meaningful information, e.g., by using "justifications"~\cite{Biran2017ExplanationAJ_survey, biran2017_human_centric_justifications, definelicht2020} to explain how decisions are made. As stakeholders' interest in intentional information was central to our study, future research should examine how this information can be conveyed meaningfully.

\subsubsection{Taking steps to connect explainability to civic education.} Our studies show that our participants were not always interested in understanding ADM systems, even though they might affect them. Assuming that democratic civil societies have the right to understand public AI systems~\cite{zuger_ai_2023} and that explanations are a means to realize this right~\cite{definelicht2020}, the question of how to redeem the interest of these stakeholders becomes essential. Research on democratic participation uses the term "civic ignorance", which is the common phenomenon that citizens are disinterested in laws and regulations shaping their lives~\cite{lau_advantages_2001, lupia_uninformed_2016}. Civic ignorance is not a stigma but an effect that applies to most people, including political experts~\cite{lupia_uninformed_2016}. \citet{lupia_uninformed_2016} describes that interest can be remedied by developing educational material that covers the "\textit{information that matters}", which is the intersection of i) information that advances the core concerns of the public and ii) information that enables the performance of a task, such as judging a political topic. 

This paper aims to accomplish this step of identifying information relevant to people affected by algorithmic decisions and sufficient to form a judgment. We argue that future work should combine these empirical insights with conceptual work on AI literacy~\cite{long_what_2020, ng_ai_2021} to design explanations that effectively convey information to these stakeholder groups. Our study outlines the first steps towards this approach: Orientating explanations along AI novices' information needs, allowing for flexible switching between different topics and explanatory stances, using guidance material such as the XAI Question Bank~\cite{liao2020} and the XAI Novice Question Bank (\autoref{fig:xainqb}), and considering to include contextual information such as intention.

\subsection{Practical implications}

\subsubsection{Resolving understanding challenges in question-driven explanations.} The direct aim of explanations is to increase a person's understanding of an ADM system~\cite{langer_what_2021}, but both designing explanations that reliably increase understanding and evaluating if they succeeded is surprisingly difficult~\cite{schmude2023}. In the following, we analyze how our question-driven explanations interacted with participant understanding and derive two concrete implications for future explanation design. 

First, \textbf{question-driven explanations succeed in letting participants independently select relevant information and guiding them to non-explored aspects of the ADM system}, for example, by providing the XAI Question Bank~\cite{liao2020} as guidance. This supports the notion of personalized explanations~\cite{conati_toward_2021, shulner-tal_enhancing_2022, martijn_knowing_2022} and could be further elaborated in a design study of a digital question-driven explanation tool based on the XAI Novice Question Bank's categories. This would allow for a better overview and more straightforward navigation of information while potentially being suitable and relevant for non-expert users. This implementation could further explore the possibility of including a generative model using an information base to generate explanations in response to users' input queries, as has already been proposed for other domains~\cite{zhao_explainability_2024}. 

While this generative explainer would allow for flexible and comprehensive explanations that could vary in detail, it also poses several challenges: First, the information used for our explanations was "scavenged"~\cite{wieringa_hey_2023} from various sources, meaning that it needs to be decided which information should be prioritized if sources are ambiguous or contradict each other. Second, non-factual information, such as explanations of intention, would require the model to reflect multiple points of view clearly and neutrally represent value conflicts likely to arise in high-risk ADM systems~\cite{raji_fallacy_2022}. Third, non-deterministic generative models are known to produce content that can vary in quality and might even contain wrong information~\cite{karinshak_working_2023}, which would be very difficult to spot for AI novices. Lastly, multiple participants in our study reported that they appreciated the social nature of the explanation format and the exchange with another person, which is in line with previous research~\cite{miller_explanation_2019} and which would be mostly lost when transferred to a generative explainer. Providing a viable implementation of generative question-driven explanations would thus require thoroughly considering these issues. 

Second, \textbf{several processes can interfere with increasing understanding through question-driven explanation, including blind spots, outsourcing, and understanding self-assessment}. We use the term \textit{blind spots} for information that participants felt was missing but that they could not acquire. This might mean a lack of "intelligibility types"~\cite{lim+dey_assessing_intelligibility2009}, i.e., a lack of guidance on which information was available and how to ask about it. Documents such as our proposed XAI Novice Question Bank (\autoref{fig:xainqb}) and the XAI Question Bank~\cite{liao2020} should help in addressing this issue by showing people the relevant categories of information. 

\textit{Outsourcing}, on the other hand, refers to sharing the process of understanding with other people to make sense of it collaboratively, a topic that seems underexplored in current XAI research. Future work should thus consider if the effectiveness of explanations can be increased by providing them to groups of people who can rely on the "expertise in other minds"~\cite{keil2006}, i.e., discuss and collaborate. An essential aspect of outsourcing understanding is the reliance on other people to aid in the understanding process~\cite{keil2006}, especially when the topic seems complex and overloaded. In our study, this prompted participants to skip parts of the understanding process entirely and instead ask the study examiner for their assessment. Considering that civil society organizations are already entrusted with providing support for people regarding finances, education, legal aid, and more, they are an intuitive point of orientation when it comes to providing explanations about high-risk ADM systems to those affected by their decisions~\cite{scott_algorithmic_2022}. We argue that future work in XAI should thus cooperate with these organizations to deploy explanations that enable AI novices to understand, negotiate, and contest decisions of ADM systems~\cite{kaun2020}.  

Lastly, question-driven explanations also require \textit{understanding self-\\assessment}, meaning people decide when their understanding is sufficiently developed. However, this can be difficult to assess~\cite{keil2006}, as perceived understanding can paradoxically decrease despite a gain in information~\cite{cheng_explaining_2019} and can also fall prone to the "illusion of explanatory depth"~\cite{rozenblit2002, chromik_i_2021}, an overestimation of one's understanding. Previous work has outlined various strategies to offset these issues, including deliberate self-explanation~\cite{duckworth_tell_2001}, better initial introductions to the system~\cite{chromik_i_2021}, and checking for people's factual understanding by asking examination questions~\cite{bucinca_proxy_2020}.  To offset the decrease in self-reported understanding stemming from the gain in new information, we propose to include an elicitation of participants' information gain in addition to their perceived understanding. A survey question could be formulated as "How much knowledge did you gain about the system?". 

\subsubsection{Matching explanations to stakeholders' needs.}  
\label{sec:discussion_stances}
For the reasons outlined, people can take very different approaches to information acquisition and understanding, depending partly on their stakeholder role and aims. As explanatory stances (mechanical, design, and intentional~\cite{dennett_intentional_1998}) describe how people make sense of information and build predictive strategies from it, they can inform explanation design by considering which stance people apply when acquiring information. While previous XAI research explored explanatory stances conceptually~\cite{Paez2019, miller_explanation_2019, zerilli_2022}, we argue that future work should investigate explanatory stances empirically for two main reasons:

First, our analysis shows that participants tended to apply different stances in their questions depending on the topic of inquiry: questions about data and system details tended to aim for technical information (mechanical), while questions about usage and context instead aimed for information on the system's practical and conceptual design aspects (design) or the values that it might enforce due to how and why it was developed (intentional). Importantly, participants can assume any of the three stances in their questions, and similar-sounding questions might aim for different information, such as "How often does the system make mistakes?". \textbf{Identifying a question's explanatory stance can thus help match the provided information to the person's information need} by considering the question's context or disambiguating which category of information it aims for. This identification can be facilitated by context information, such as the participants' previous questions, or by explicitly inquiring how the information is relevant to the participant.

Second, \textbf{encouraging the use of different explanatory stances can support a change in perspective}. Work in the cognitive sciences, for example, describes the possibility that the design stance might be preferred over the mechanical stance by default~\cite{lombrozo_mechanistic_2019}.\footnote{Preference in explanatory stances is a controversial discussion~\cite{keil_2021_mechanistic_explanation, lombrozo_mechanistic_2019, kelemen_professional_2013}.} In our study, we observed that participants used the design and intentional stance instead of the mechanical stance, especially before receiving the XAI Question Bank~\cite{liao2020}. However, the boundaries might not be as clear-cut due to the interpretation required when assigning explanatory stances to participants' questions. Still, considering which information might prompt participants to take a specific stance could be useful for the design of explanations, for example, by providing material that prompts a variety of stances and steers attention to information that is not explored autonomously. For example, introducing the XAI Question Bank~\cite{liao2020} in the second inquiry phase of our study encouraged participants to ask questions aiming for technical details (mechanical), which they reported to be very useful as they had not considered these perspectives before.  

\subsection{Summary of implications}
\label{sec:discussion_implications}

In this section, we briefly summarize the implications that our findings and discussion have for the design of explanations.

\textbf{Theoretical}
\begin{enumerate}
    \item AI novices' information needs vary by use case. The proposed XAI Novice Question Bank (\autoref{fig:xainqb}) expands the XAI Question Bank~\cite{liao2020} with questions on system context and usage, highlighting relevant information for novices. Future work should explore integrating this information into formats that can be flexibly adapted to different systems.
    \item Information on \textit{why} an ADM system is deployed can be relevant for various stakeholders. Typical explanations focus on the system's workings but not on this contextual information. XAI research should consider how to best convey information on the intention behind a system's development and deployment.
    \item Researchers developing explanations for those affected by algorithmic decisions should note that people might be disinterested in explanations due to a perceived inability to understand ADM or feelings of powerlessness. Future research should further examine how these factors impact explanations for groups who are affected by ADM systems.
\end{enumerate}

\newpage

\textbf{Practical}
\begin{enumerate}
    \item Question-driven explanations let stakeholders acquire information until their understanding meets their aims. A digital implementation with a generative explainer model could apply these explanations practically. Challenges to understanding can be addressed by guiding available information, considering if understanding can be outsourced to others, and eliciting information gain in addition to perceived understanding.
    \item Information can be understood from different angles, and some may be more intuitive for specific stakeholders: domain experts might ask about a system's design aspects. In contrast, job-seekers might ask about the beliefs governing its deployment. Explanatory stances can help to identify which angle someone will take, allowing for better matching of explanations to their needs. Encouraging stance-switching, such as providing technical information to AI novices, can lead to valuable perspective changes.
\end{enumerate}

\section{Limitations}
\label{sec:limitations}


Our participants were recruited from a local job agency and an apartment complex from the same geographical region, perhaps resulting in regional or cultural biases. Due to the limited sample size, we did not analyze the impact of sex and/or gender on our results. While this limits the generalizability of our results regarding sex and/or gender aspects, it does not limit the overall validity of our results. Further, the articulated information needs and observed processes of understanding are tied to the presented use cases, as demonstrated in Section~\ref{sec:rq2}, and to the verbal exchange with the study examiner, meaning that participants would likely ask different questions if the information was presented in another fashion or by another person. Concerning the size of the participant sample, we are guided by research on qualitative methods, which suggests that 16 to 24 interviews are needed to reach adequate code and meaning saturation~\cite{hennink_code_2017}. 

\section{Conclusion}
\label{sec:conclusion}

The use of public AI systems in society can have a considerable impact on people affected by algorithmic decisions. These stakeholders should be able to understand AI systems in order to take action, such as contesting decisions or asking for human intervention. Explanations can support this empowerment, but the information needs of this stakeholder group---who tend to be AI novices---have rarely been covered in explanation approaches. To address this gap, we provide a collection of information needs titled the "XAI Novice Question Bank" (\autoref{fig:xainqb}), compiled from interviews with 24 participants and covering two ADM use cases (employment prediction and a health wristband). Our approach aims to extend the concept of question-driven explanations, inspired by the XAI Question Bank~\cite{liao2020}, to lay audiences and highlights the relevance of information needs about an ADM system's context and usage.
Further, we analyze participant understanding by applying cognitive theories to examine from which perspective participants approach information acquisition. We provide suggestions on matching explanations to how stakeholders make sense of information and strategies to overcome understanding challenges. Lastly, we examine participants' perceptions of the systems' risk and benefits, finding that higher perceived risk likely prompts participants to inquire about intentional information, such as the motivation and consequences of deploying a system. We close with five critical implications that our findings have for the design of explanations for AI novices affected by algorithmic decisions. More research is needed to create suitable explanations for this crucial stakeholder group. 



\begin{acks}
This work has been funded by the Vienna Science and Technology Fund (WWTF) [10.47379/ICT20058] as well as [10.47379/ICT20065]. We further thank the Vienna Job-TransFair team for their valuable contributions. 
\end{acks}

\bibliographystyle{ACM-Reference-Format}
\bibliography{references.bib}

\newpage

\appendix
\def\thesection{\Alph{section}}

\section{Supplementary Material}

In the following, we provide additional background information on the \amsalgorithm (Section~\ref{sec:suppmat_ams_algorithm}), the reduction process of the XAI Question Bank~\cite{liao2020} (Section~\ref{sec:suppmat_xaiqnb_reduction}), and examples of how unavailable information was handled in the interviews (Section~\ref{sec:suppmat_information_unvailable}). 

\section{Additional information about the \amsalgorithm}
\label{sec:suppmat_ams_algorithm}

\subsection{Development and public discourse}

The AMS algorithm incited public discourse about the benefits and risks of its deployment. According to the documentation accompanying its development~\cite{Holl2018, Holl2018_Standards}, the algorithm’s prediction were meant to support personnel consultants in the Public Employment Agency, under the condition that the job-seeker would be given a voice in the discussion about their employability. Further, discrepancies between algorithmic and human estimations of employability were meant to be reinserted into the algorithm to improve predictions, and finally, the predictions made by the algorithm were only to be used in this specific instance of job-seeker support and in no other capacity. According to the Agency’s communication to the public, the deployment pursued three concrete overarching goals: a) increase in the efficiency of consultation, b) increase in the effectiveness of support measures, and c) standardisation of support measures and prevention of arbitrariness~\cite{allhutter_bericht_ams-algorithmus_2020}.

However, reports on the \amsalgorithm describe the risks of problematic adaptation processes~\cite{lopez_reinforcing_2019, allhutter_bericht_ams-algorithmus_2020}, including: overreliance on the statistical analysis, insufficient training of staff regarding overruling the algorithm’s  suggestions, self-fulfilling prophecies of personal attributes that are controlled not by the individual but by their social environment, and pre-adjustment to the algorithmic suggestion by the job-seeker in order to avoid bad classifications. Transparency and ongoing scrutiny of the algorithm are listed as necessary measures to prevent these risks.~\citet{allhutter_bericht_ams-algorithmus_2020} further expanded on these points, listing additional risks of the algorithm's deployment: Reinforcement of societal biases due to reliance on historical data without adaptation to social change; systemic discrimination due to the assignment of vulnerable groups to certain constellations; and a shift from individual consultation to an overuse of computer-assisted decisions and automating of interpersonal tasks.

\subsection{Features used by the AMS Algorithm}

\begin{table}[H]
\caption{The \amsalgorithm used a small set of features to calculate employability scores. The features as described below were reported by~\citet{Allhutter2020en}. The prior occupational career is constituted by four subvariables. "Cases" describe the number of times a job-seeker registered at the employment agency, "intervals" refers to a pre-defined time range. "Measures" describe support measures such as qualification courses and subsidization. For a description and examples of how these features are weighted, please refer to~\citet{Holl2018}.}
\begin{tabular}{p{0.35\textwidth}|p{0.55\textwidth}}
\hline
Variable & Nominal values \\ \hline
Gender & Male/Female \\
Age group & 0–29/30–49/50+ \\
Citizenship & Austria/EU except Austria/Non-EU \\
Highest level of education & Grade school/apprenticeship, vocational school/high- or secondary school,   university \\
Health impairment & Yes/No \\
Obligations of care (only women) & Yes/No \\
Occupational group & Production sector/service sector \\
Regional labor market & Five categories for employment prospects in assigned AMS job center \\ \hline
Prior occupational career & [Subvariables as described below] \\ \hline
Days of gainful employment within 4 years & $<$75\%/$\geq$75\% \\
Cases within four 1 year intervals & 0 cases/1 case/min. 1 case in   2 intervals/min. 1 case in 3 or 4 intervals \\
Cases with duration longer than 180 days & 0 cases/min. 1 case \\
Measures claimed & 0/min. 1 supportive/min. 1 educational/min. 1 subsidized employment
\end{tabular}
\label{fig:ams_feature_table}
\end{table}

\newpage
\section{Reduction of the XAI Question Bank}
\label{sec:suppmat_xaiqnb_reduction}

\begin{longtable}{p{0.17\textwidth}|p{0.4\textwidth}p{0.32\textwidth}}
\caption[Reduction of XAI Question Bank]{In the second inquiry phase (described in Section 3 in the main text), participants were provided with a reduced version of the XAI Question Bank~\cite{liao2020}. The amount of questions compiled in the XAI Question Bank~\cite{liao2020} would have made its application in the study impractical due to the time necessary to scan and understand all items. Unsuitable questions were thus omitted and similar questions summarized, information for omitted questions was then given in response to similar questions. The table below describes each question in the XAI Question Bank and a brief description of which question was included or omitted in the reduced version.}\\
Category & Question & Comment \\ \hline
Input & What kind of data does the system learn from? & Included \\
 & What is the source of the data? & Included \\
 & How were the labels/ground-truth produced? & Omitted: Question would   require conceptual explanation\\
 & What is the sample size? & Included \\
 & What data is the system NOT using? & Omitted: Information was given   in response to question about data in general \\
 & What are the limitations/biases of the data? & Included \\
 & How much data {[}like this{]} is the system trained on? & Omitted: Information was given   in addition when asked about sample size \\ \hline
Output & What kind of output does the system give? & Included \\
 & What does the system output mean? & Included \\
 & How can I best utilize the output of the system? & Omitted: Participants where not   meant to assume the role of users \\
 & What is the scope of the system's capability? Can it do. ..? & Included \\
 & How is the output used for other system component(s)? & Included \\ \hline
Performance & How accurate/precise/reliable are the predictions? & Included \\
 & How often does the system make mistakes? & Included \\
 & In what situations is the system likely to be correct/incorrect? & Omitted: Information was given   when asked about system's likely kind of mistakes \\
 & What are the limitations of the system? & Included \\
 & What kind of mistakes is the system likely to make? & Included \\
 & Is the system's performance good enough for. . . & Omitted: Requires specific knowledge about domain tasks \\ \hline
How (global) & How does the system make predictions? & Included \\
 & What features does the system consider? & Included \\
 & Is {[}feature X{]} used or not used for the predictions? & Omitted: Information given when   asked about features in general \\
 & What is the system's overall logic? & Included \\
 & How does it weigh different features? & Omitted: Information given when   asked about features in general \\
 & What rules does it use? & Omitted: Information given when   asked about logic \\
 & How does {[}feature X{]} impact its predictions? & Omitted: Information given when   asked about features in general \\
 & What are the top rules/features it uses? & Omitted: Does not generally   apply to use cases \\
 & What kind of algorithm is used? & Included \\
 & How are the parameters set? & Omitted: Information given when   asked about features in general \\ \hline
Why & Why/how is this instance given this prediction? & Omitted: Specific instances of predictions were not provided, information was given when conversation turned to examples \\
 & What feature(s) of this instance leads to the system’s prediction? & Same comment applies \\
 & Why are {[}instance A and B{]} given the same prediction? & Same comment applies \\ \hline
Why not & Why/how is this instance NOT predicted...? & Same comment applies \\
 & Why is this instance predicted P instead of Q? & Same comment applies \\
 & Why are {[}instance A and B{]} given different predictions? &  Same comment applies \\  \hline
What If & What would the system predict if this instance changes to...? & Same comment applies \\
 & What would the system predict if this feature of the instance changes   to...? & Same comment applies \\
 & What would the system predict for {[}a different instance{]}? & Same comment applies \\ \hline
How to be that & How should this instance change to get a different prediction? & Same comment applies \\
 & How should this feature change for this instance to get a different   prediction? & Same comment applies \\
 & What kind of instance gets a different prediction? & Same comment applies \\ \hline
How to still be this & What is the scope of change permitted to still get the same prediction? & Same comment applies \\
 & What is the {[}highest/lowest/…{]} feature(s) one can have to still get the   same prediction? & Same comment applies \\
 & What is the necessary feature(s) present or absent to guarantee this   prediction? & Same comment applies \\
 & What kind of instance gets this prediction? & Same comment applies \\ \\ \hline
Others & How/what/why will the system change/adapt/improve/drift over time?   (change) & Included \\
 & How to improve the system? (change) & Included \\
 & Why using or not using this feature/rule/data? (follow-up) & Omitted: Information was provided when asked about features in general \\
 & What does {[}ML terminology{]} mean? (terminological) & Included \\
 & What are the results of other people using the system? (social) & Included \\ \hline
\end{longtable}

\newpage

\section{Example transcripts for re-formulation of interview inquiries}
\label{sec:suppmat_information_unvailable}

When questions could not be answered due to a lack of information in the inquiry phases, they were re-formulated by the study examiner with the intention of asking participants why the requested information would be relevant to them. In the following, we present two excerpts from interview transcripts for illustration. 

In the first excerpt, participant A12 asks about the inclusion of an IQ parameter as an additional means of classification:

\begin{quote}
    \textit{A12: Could the system theoretically also classify people according to the IQ [...]? \\
    Examiner: Theoretically, this could be included. Why would you like to know? \\
    A12: Well, that's what I was wondering. But according to the details of the algorithm, it's also difficult to assign such an IQ, I think? It could only be determined by education.}    
\end{quote}

In the second excerpt, participant B3 inquires about functionality of the health wristband, describing it to be both relevant for patient and care-giver and mentioning notions of data privacy:

\begin{quote}
    \textit{B3: Can this be expanded to show blood pressure and diabetes, if diabetes is high? [...] \\
    Examiner: Would that be for the person wearing the bracelet or for the nursing staff? \\
    B3: Could be for both. The person should know how they look and the caregiver too, because they [the patients] like to fib. \\
    Examiner: Would you be interested in that? If we assume that we have this data now. \\
    B3: Sure. [...] Of course I would be interested to know what my blood pressure looks like. [...] Someone else can always see my data and I can't, I don't think that's okay.}
\end{quote}


\end{document}